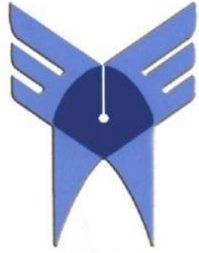

Islamic Azad University
Tehran Science and Research (East Azarbaijan Branch)

# A Thesis Submitted in Partial Fulfillment of the Requirements for the Master of Science Degree in Computer Engineering – Computer System Architecture

**Subject:**
Lightweight Hybrid Block-Stream Cryptographic Algorithm for the Internet of Things[1]

**Thesis SEPARvisor:**
Mir Kamal Mirnia Herikandi (Ph. D.)

**Thesis Advisor:**
-

**By:**
Arsalan Vahi

Spring 2020
Technical and Engineering Department

---

[1] This document has been translated from a Persian thesis.

The contents of this thesis are based on the article "*SEPAR: A New Lightweight Hybrid Encryption Algorithm with a Novel Design Approach for IoT,* which was previously published in "*Wireless Personal communication*" Journal. The article represents a part of the research work conducted for this thesis.

## Abstract

In this thesis, a novel lightweight hybrid encryption algorithm named **SEPAR** is proposed, featuring a 16-bit block length and a 128-bit initialization vector. The algorithm is designed specifically for application in Internet of Things (IoT) technology devices. The design concept of this algorithm is based on the integration of a pseudo-random permutation function and a pseudo-random generator function. This intelligent combination not only enhances the algorithm's resistance against cryptographic attacks but also improves its processing speed. The security analyses conducted on the algorithm, along with the results of NIST statistical tests, confirm its robustness against most common and advanced cryptographic attacks, including linear and differential attacks. The proposed algorithm has been implemented on various software platform architectures. The software implementation was carried out on three platforms: 8-bit, 16-bit, and 32-bit architectures. A comparative analysis with the **BORON** algorithm on a 32-bit ARM processor indicates a performance improvement of **42.25%**. Furthermore, implementation results on 8-bit and 16-bit microcontrollers demonstrate performance improvements of **87.91%** and **98.01%** respectively, compared to the **PRESENT** cipher.

# Table of Contents

## Chapter 1: Introduction to the Internet of Things and Study Background

## Chapter 2: Literature Review on Lightweight Encryption Algorithm Design

# List of Figures

# List of Pseudocodes

# List of Tables

k

# Chapter 1: Introduction to the Internet of Things and Research Background

**Objectives of the Chapter:**

1. Introduction to the Internet of Things (IoT)

2. Challenges in Securing IoT

3. Research Background

4. Thesis Structure

# Introduction

In this chapter, the concept of the Internet of Things (IoT) and the relevant research context is introduced. Initially, it provides a definition of IoT technology, followed by a discussion of its major challenges. The chapter concludes by outlining the structure of the thesis.

## 1.1 Introduction to the Internet of Things

Advancements in embedded system manufacturing and the growing tendency toward their widespread applications, alongside the development of internet communication infrastructure—particularly driven by the popularity of wireless communication technologies over the past decade—have enabled the vision of a more convenient life through the integration and smartening of everyday objects. This concept, first proposed by Kevin Ashton in 1999, gave birth to the concept of "Internet of Things" technology [1]. IoT has emerged from the convergence of five mature technologies: ubiquitous computing, sensor networks, big data, cloud computing, and radio frequency identification (RFID) technology [2]. Key security challenges in IoT include privacy protection, scalability, and the heterogeneous structure of its networks [3]. The main goal of IoT is to integrate objects into human daily life, allowing them to participate in routine activities. However, such integration inevitably exposes personal and confidential data to surrounding smart objects. Without adequate security considerations, applying IoT to everyday objects is not practically feasible and may lead to harmful consequences rather than comfort and convenience.

IoT refers to a collection of objects within an environment that are interconnected through a dynamic public network, capable of self-configuration and communication based on standard and universally accepted protocols. These objects collect data from the environment, transmit it via the internet to a data analysis center, and receive specialized intelligent services in return [4]. Figure 1-1 presents the fundamental structure of the system. Objects within the environment—such as household items, factory equipment, street infrastructure, clothing, shoes, and balls—that need to be internet-connected, are linked to intermediary units (electronic boards and RFID tags), which translate internet-based instructions and messages into forms understandable by the objects. Each intermediary unit connects to both an object and either the internet or a local network. Depending on their internal architecture, these units may feature Wi-Fi, USB, or other interfaces, allowing them to connect to access points and obtain unique IP addresses. This connectivity facilitates cooperation and interaction among smart objects. These units may also include internal web servers or connect to external professional web servers via the internet. Alternatively, the systems may be managed through a centralized control center on the internet, delivering services such as smart homes, smart transportation, and more, based on preconfigured plans [5].

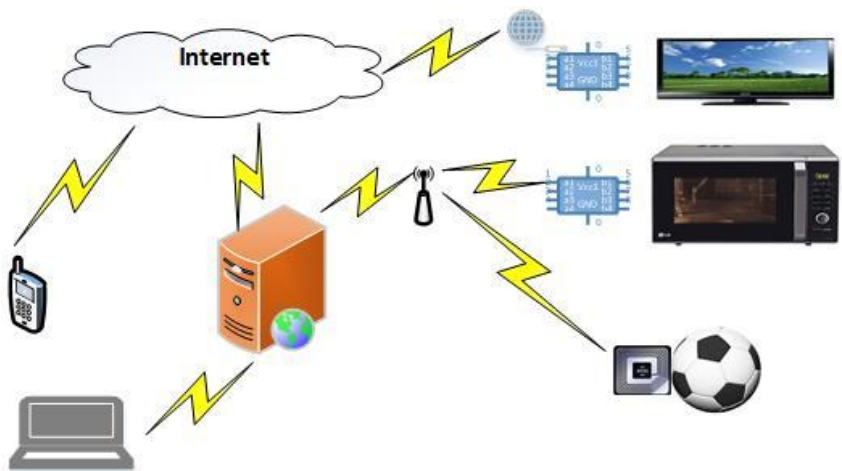

**Figure 1-1: IoT System Structure**

The core and inherent characteristics of IoT systems are primarily based on ubiquitous computing and embedded systems. Typically, IoT systems exhibit the following features [6]:

• **Presence of Smart Objects**: Intelligence in objects arises from the integration of algorithms and computation (software and hardware). These objects must be capable of performing basic processing tasks to determine whether the environmental data they collect is significant enough to be sent to the cloud for service provision, and to make appropriate decisions.

• **Object Connectivity**: In a dynamic, scalable network of heterogeneous mobile objects, each object must be capable of communication and compatibility with others. Machine-to-machine (M2M) communication protocols are used to facilitate their interoperability within the network, enabling both data production and consumption.

- **Energy Constraints**: Most objects have limited resources and energy. They are typically equipped with low-cost, low-performance processors and possess minimal storage capacity. Moreover, many of these objects rely on batteries as their energy source, limiting the feasibility of employing complex algorithms.

- **Limited Security**: Due to the dynamic network structure and mobility of objects, they may enter uncontrolled environments lacking oversight, increasing their vulnerability to various threats. Sensors are often installed in publicly accessible locations, making them physically susceptible to theft and damage. Securing the objects, network, and data flow is essential for ensuring overall system security.

IoT aims to connect all potentially connectable elements capable of interacting via the Internet, thereby enabling a more secure and convenient lifestyle. Some common IoT applications include [2]:

- **Energy and Power Management**: Systems connected to the internet that use sensors to monitor and reduce energy consumption. Examples include smart lamps and controlled ovens.

- **Home and Building Automation**: IoT can function as a remote controller and monitor for household systems. Examples include phones, televisions, and similar appliances.

- **Transportation Systems**: In urban environments, IoT enables automated systems such as adjustable traffic lights and smart surveillance cameras that report less congested routes and help reroute traffic efficiently.

Given the intrinsic features of IoT outlined above, it is necessary to first study the types of potential attacks in this environment to propose an effective security solution. This is addressed in the following section.

## 1.2 Security Challenges on the Intenet of Things

The previous section provided an overview of the IoT concept and its applications. This section highlights common types of attacks that are crucial for security engineers to analyze. The following are notable examples:

- **Authentication Attacks**: A prevalent authentication method in IoT involves attaching tags containing identity information to objects, which are then read using a tag reader. If an attacker gains access to these tags and their information, they can read or modify the data, or even duplicate and transfer it to another tag, thereby creating a forged identity [7].

- **Eavesdropping**: One of the simplest and most dangerous passive attacks, eavesdropping targets user privacy. Due to the wireless nature of communication among objects, an attacker can intercept the communication channel to listen in on transmitted information. This becomes especially serious if sensitive data such as passwords or personal details are exchanged without encryption [6].

- **Spoofing**: If an attacker injects false data into the system and the system mistakenly accepts it as originating from a legitimate source, the attacker can exploit this vulnerability to gain access.

- **Resource Drain Attacks**: In this type of attack [8], a malicious object attempts to exhaust the battery power of other objects by forcing them to perform unnecessary operations.

- **Denial-of-Service (DoS) Attacks**: This attack involves the attacker sending abnormal traffic to a target object, thereby preventing legitimate network traffic from reaching that object.

Based on the characteristics and attack types described above, two primary categories of security requirements in the IoT context can be defined [9]:

1. **Network Security**: Essential services in this domain include authentication, confidentiality, data integrity, and availability.

2. **Privacy and Identity**: This is one of the most challenging aspects due to the presence of individuals and the vast amount of data collected by the IoT system. The multitude of connected objects and the complex relationships among devices, services, owners, and users make privacy and identity protection particularly critical. Object owners generally do not wish to be tracked merely due to the presence of smart devices in their environments.

Meeting these security needs requires the development of appropriate security solutions, among which cryptographic techniques are considered a promising approach.

## 1.3 Study Background

The development of an effective cryptographic algorithm can serve as a robust security solution to provide essential security services within IoT. Table 1-1 presents a summary of proposed security solutions along with the layered architecture of the Internet of Things [10].

Table 1-1: Overview of Proposed Security Solutions Aligned with IoT Layered Architecture

| IoT Layered Architecture | Elements | Solutions |
|---|---|---|
| Application Layer | Personlized Information Services<br>Smart Transportation Services<br>Monitoring System services | Identity agreement<br>Key Agreement<br>Protecting Privacy<br>Encryption Mechanism |
| **Data Processing Layer** | Cloud Processing<br>Intelligence Processing | Intrusion Detection<br>Secure Cloud Processing<br>Antivirus and Firewalls |
| **Network Layer** | Mobile Internet Communation Netwrok<br>Satelite Networks<br>Communication Protocols | Encryption Mechanism<br><br>Securing Commnication |
| **Perception Layer** | Radio Frequency Tag Readers<br>Sensors | Simple Encryption<br>Key Agreement<br>Protecting Sensitive Data |

The objective is to develop a cryptographic algorithm that meets the necessary security requirements on the Internet of Things and is designed to be multipurpose, aligning with the characteristics of IoT devices as previously discussed. In Chapter 2, lightweight cryptographic algorithms will be briefly introduced.

## 1.4 Thesis Structure

This thesis is organized into four chapters. Chapter 1 provided a general definition of the Internet of Things and presented its key security challenges. Chapter 2 offers a comparative analysis of block and stream cipher methods, highlighting their respective features, along with a review of commonly used symmetric and public-key encryption algorithms. Chapter 3 introduces a novel cryptographic algorithm that combines concepts from both block and stream ciphering techniques. Finally, Chapter 4 presents conclusions and outlines future research directions.

**Summary and Conclusion**

This chapter introduced the concept of the Internet of Things and the associated security challenges. Among these challenges, the development of a new lightweight cryptographic algorithm was identified as the focus of this research, and its significance was discussed. Therefore, to provide a deeper understanding, the next chapter will compare block and stream ciphering methods and explore the various features employed in existing conventional algorithms.

# Chapter 2: Literature Review on Lightweight Encryption Algorithm Design

## Chapter Objectives:

- Survey of Design Methodology for Lightweight Ciphers

- Analysis of Key Features in Various Cryptographic Methods

- Comparative Study of Existing Design Methods

- Proposal of a Novel Cryptographic Algorithm

# Introduction

This chapter begins with a definition of cryptography and an examination of symmetric and asymmetric cryptosystems. Given the compatibility of the Internet of Things (IoT) environment with symmetric cryptosystems, the two primary encryption methods of block cipher and stream cipher, are explored. A comparison of the advantages and disadvantages, along with an analysis of the features of these two methods, leads us to explore a balance between them and investigate previously designed lightweight cryptographic algorithms, ultimately guiding us toward the design of our own hybrid algorithm called **SEPAR**.

## 2.1 Cryptography

The term cryptography is derived from two Greek words: "crypto," meaning hidden, and "graph," meaning writing, and refers to the science of hidden information. One of the famous historical ciphers is the Caesar cipher, which was used by Julius Caesar to prevent the leakage of military information by shifting each letter by three units, thus preventing his messages from being realized by enemies. Since 1949, with the historical review by Claude Shannon, a new window was opened to modern cryptography, and competition for utilizing cryptographic capabilities in various fields, including telecommunications and information technology, began [11]. Four essential requirements for protecting systems and data against adversaries are outlined as follows [12]:

- **Confidentiality Requirement**: Only the sender and receiver of the communication should be able to understand the contents of the transmitted data. This requirement is achievable only through cryptographic solutions.

- **Data Integrity Requirement**: Ensures that data received in the receiving part is original and has not been altered in any way, maintaining exactly what was sent by the original sender. If the content of the communication has been modified, it must be understandable by the parties involved.

- **Authentication Requirement**: The sender and receiver must be able to verify each other's identities. Any fraudulent actor must be identifiable and understandable.

- **Non-repudiation Requirement**: Ensures that no entity can deny having performed an action previously. For instance, in network communications, the sender of a message should not be able to deny sending it.

Among these four services, **confidentiality** is the core service, and all cryptographic algorithms must provide it, whereas the provision of other services is optional.

## 2.2 Lightweight Cryptography

The 1970s marked the emergence of cryptography in both scientific and commercial spaces, with its widespread application in everyday operations. Alongside these advancements, the push for creating smaller electrical and telecommunications components with lower energy consumption gave rise to a new branch in cryptography known as **lightweight cryptography**. The main challenge in designing lightweight ciphers is striking a balance between three important parameters: security, cost, and efficiency. The security of a cryptosystem is measured by the length of its key, while the performance of the cryptographic algorithm depends on its throughput. To determine the cost, factors such as space, memory consumption, and code size are examined. It is important to note that in defining lightweight cryptography, no specific boundaries are set for each of these three parameters—cost, security, and efficiency—as they largely depend on the environment in which the system is deployed.

## 2.3 Symmetric and Asymmetric Cryptosystems and Their Characteristics

Cryptosystems are categorized into symmetric and asymmetric systems based on the number of keys used for encryption operations. Symmetric systems, also known as **secret key cryptography**, use a single shared key for both encryption and decryption. Figure 2.1 illustrates the working mechanism of this cryptosystem.

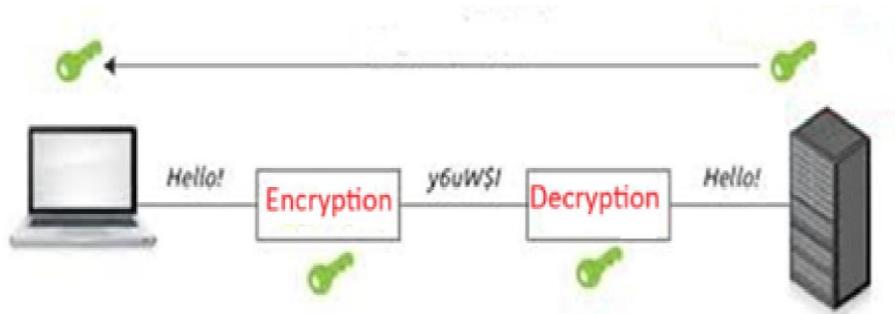

**Fig 2.1: Symmetric Encryption**

A fundamental requirement in the application of these cryptosystems is the ability to maintain the secrecy of the key during its exchange between the sender and the receiver. Additionally, when using these cryptosystems, employing a strong cryptographic algorithm is critical. By strong, it is meant that an adversary, having access to the encryption algorithm and the ciphertext, should not be able to reverse the operation (i.e., decrypt the message) and retrieve the key and original message.

One of the challenges with symmetric encryption algorithms is the challenge of key distribution between authorized parties, maintaining the confidentiality of the key, and the inability to provide authentication and integrity services independently or without combining with other algorithms. In contrast to these practical weaknesses, efficient hardware and software implementations of these cryptosystems have made them a viable option for applications. Symmetric encryption algorithms are classified into two categories: block ciphers and stream ciphers. By combining these two methods, a new category, known as the **hybrid category**, can also be introduced.

In the second category, asymmetric cryptosystems are placed. The concept of encryption in these systems is based on the idea that the algorithm does not decrypt what it encrypts with the same key; decryption is performed with a different key. Figure 2.2 shows the operation of this cryptosystem.

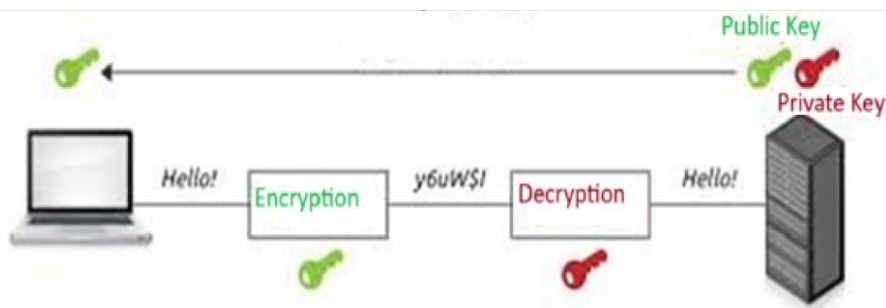

**Fig 2.2: Asymmetric Encryption**

One of the two distinct keys in this cryptography is the **public key**, which is accessible to everyone, and the other is the **private key**, which is kept confidential by the user. In this method, anyone can encrypt a message with a public key, while decrypting the encrypted message is only possible with the private key associated with that public key. Although theoretically it is possible to obtain the private key from the public key, in practice, obtaining and computing requires considerable time and computational resources. Asymmetric key algorithms are significantly more secure compared to symmetric key algorithms; however, their implementation is much more complex, resulting in a much slower execution speed.

## 2.3.1 Comparison of Symmetric and Asymmetric Encyptions

Each of the symmetric and asymmetric cryptosystems, depending on their characteristics, has its own advantages and disadvantages. Table 2.1 summarizes the advantages and disadvantages of these systems.

Table 2.1: Comparison of Symmetric and Asymmetric Cryptosystem

| Feature | Symmetric Encryption | Asymmetric Encryption |
|---|---|---|
| What must remain secret and what is public | Secret: key<br>Public: plaintext and algorithm | Secret: private key<br>Public: plaintext and algorithm |
| Key length | Typically short | Typically long |
| Changing keys in communication | Both parties must agree and change the key | Each person keeps their private key; public key is freely distributable |
| Key Storage | Must be Securly Stored | Private key must be securely stored |
| Ciphertext Calculation | Using two basic principles: confusion and diffusion | Using complex mathematical calculations |
| Implementation and design aspects | Easier to implement | Due to complex mathematical operations, asymmetric encryption typically demands significant computational resources. For example, if implemented on an 8-bit micro with limited resources, their performance is 100 or 1000 times slower than symmetric algorithms. |
| Execution Speed | Relatively 1000 times faster. | Complex mathematical calculations make them slower. |
| Important Applications | Very Efficient for Encryption | Due to the high amount of cryptographic calculations, these schemes are not mainly used for cryptographic operations but rather are mostly used in applications such as email that require authentication and digital signatures. They are also mostly used to establish an exchange key for symmetric cryptographic algorithms between communicating parties [13]. |

Based on the comparison in Table 2.1, it can be concluded that symmetric encryption is more suitable for resource-limited devices in the IoT environment. The next section will examine block and stream encryption methods and study their features.

## 2.4 Bock and stream Cipher Techniques: Features and Evaluations

Since symmetric cryptosystems are suitable for resource-limited environments, the structure of algorithms used in these cryptosystems will be reviewed next. These algorithms are classified into two broad categories: block ciphers and stream ciphers.

### 2.4.1 Block Ciphers

Block ciphers begin the encryption operation by processing blocks of data. The basic function of these algorithms is based on a pseudo-random permutation, where the internal structure of the algorithm, along with the key, determines which input block is mapped to which output block.

#### 2.4.1.1 Core Components Used in Block Algorithms

A block cipher algorithm is composed of four basic elements: key expansion, key mixing, internal structure, and round function [14]. The general structure of a block cipher algorithm is shown in Figure 2.3.

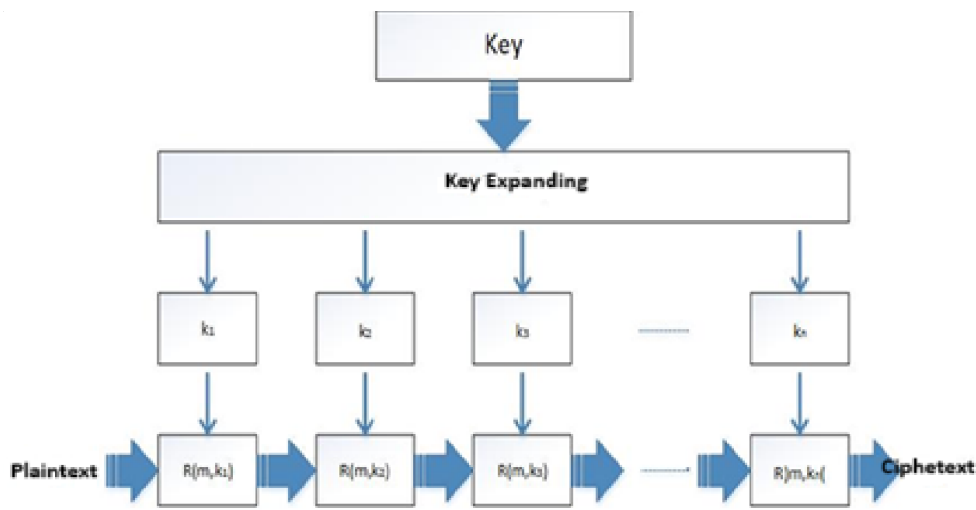

**Fig2.3: Overall Schematic for a block cipher algorithm**

The **key expansion element** is responsible for dividing the key input by the user into several equal-sized subkeys within the algorithm. Its structure is designed as a good pseudo-random generator, although its complexity should be much lower compared to the algorithm itself. This part can be viewed as a small algorithm that does not need to be reversible. By using a suitable generator, the analysis of subkey generation becomes difficult [14].

Block cipher round functions are divided into two categories: **confusion functions** and **diffusion functions**. Diffusion functions are linear functions that prevent leakage of any statistical information from the plaintext into the ciphertext. These functions, using linear operators, uniformly distribute the input across the output space. This property causes the influence of each bit of the plaintext or key to be spread out and minimized in the ciphertext. The goal of these functions is to create a lack of correlation between the bits of the input text, key, and output text, so that statistical relationships are eliminated, and cryptanalysis becomes more complex. One of the most commonly used elements in algorithm design to implement this feature is the **permutation box**. This element only serves the purpose of swapping the input bits, thereby achieving the diffusion property. It is a reversible function, so it can be used again to reverse the message to its previous form. This element is highly suitable for hardware implementation, as it requires no gates and is made solely of wiring. In contrast, the complexity of its software implementation arises from the need for heavy bitwise operations.

**Confusion functions** are non-linear functions that prevent the disclosure of any key information through the analysis of plaintext and ciphertext. These functions, in addition to uniformly distributing the input across the output space, increase the complexity of the encryption and its resistance to attacks such as **linear** and **differential cryptanalysis** due to their non-linearity. The **substitution box** is one of these confusion elements that maps **n** bits of input to **m** bits of output. S-boxes are typically static lookup tables used to provide non-linearity and create a complex Boolean relationship between the plaintext and ciphertext. S-box are suitable for software implementation because they can be stored in small memory. It should be noted that security can only be achieved by applying confusion properties. A **single-key lookup table** with 64-bit input and 64-bit output is strong enough, but because implementing this large lookup table is costly, most elements that apply confusion are combined with elements that apply diffusion so that confusion elements can be implemented with smaller lookup tables.

In the overall algorithm structure, there are different arrangements for confusion and diffusion functions. Two common structures, **SPN (Substitution-Permutation Network)** and **Feistel**, are illustrated in Figure 2.4 [12].

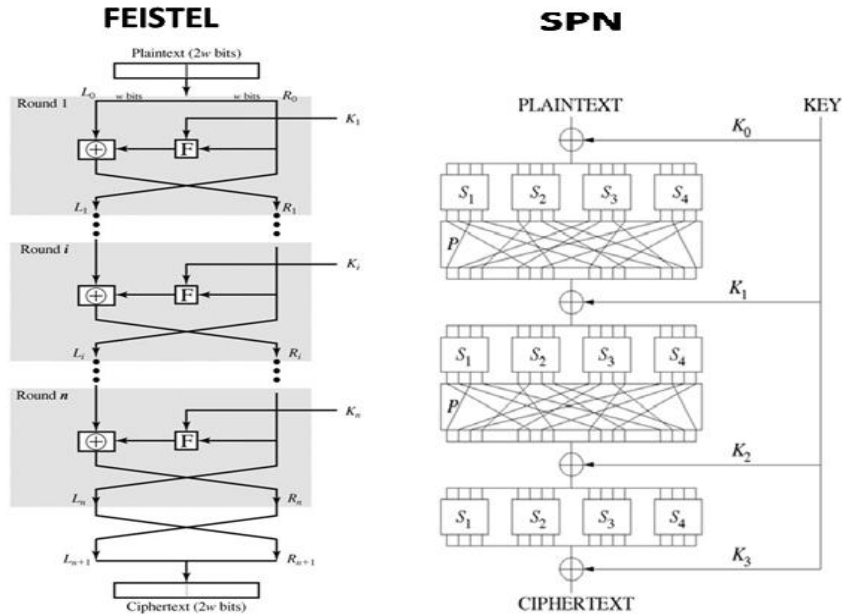

**Fig 2.4: SPN and Feistel Structures**

In the SPN (Substitution-Permutation Network) structure, the encryption operation is based on two actions: substitution and permutation. The combination of these two operations in a chained fashion adds sufficient complexity to the algorithm. In the Feistel structure, the input block is split into two halves: the left and the right. In each round, one half remains unchanged, while the other half undergoes a highly non-linear transformation through a function. Afterward, the two halves are swapped, and this process is repeated for multiple rounds. The strength of the encryption algorithm in the Feistel structure lies in the properties of this non-linear function. The key feature of the Feistel structure is that the encryption and decryption algorithms are identical, eliminating the need to invert the round function [15].

### 2.4.2 Stream Ciphers

A stream cipher, unlike block ciphers, operates by encrypting one bit at a time as it is received. The initial idea behind this encryption method is derived from the One-Time Pad (OTP). OTP is an encryption algorithm in which the plaintext is encrypted using a random key sequence equal to or longer than the message length. It has been proven that this encryption method is theoretically unbreakable, provided that the following four conditions are met [16]:

1. The length of the key used for encryption must be at least equal to or longer than the message being encrypted.

2. The key must be truly random, not generated by a simple computational function.

3. The key must be used only once; after its use, it should be immediately destroyed, both by the sender and the receiver.

4. There should only be two copies of the key: one for the sender and one for the receiver.

The advantage of this algorithm is its high speed in hardware, and its implementation is relatively simple. However, the primary challenge lies in the key size, which must always be at least as long as the plaintext. This raises issues related to key management and storage [14]. Figure 2.5 provides a representation of the structure of these algorithms [17].

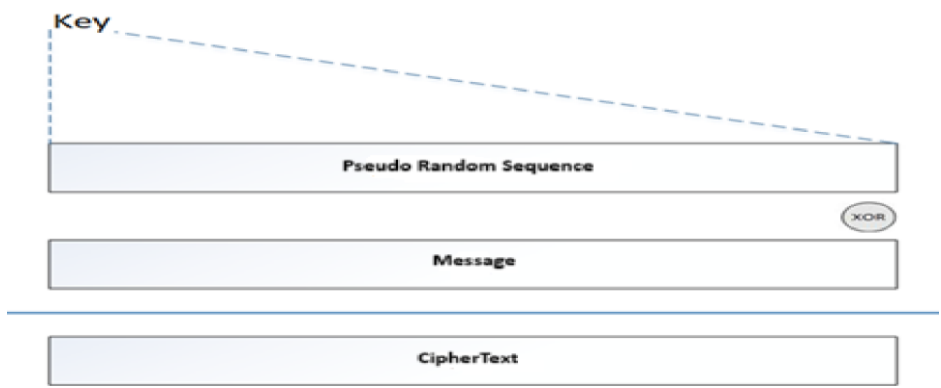

**Fig 2.5: Overall Structure for Stream Ciphers**

Each stream cipher requires three stages for operation:
1. **Initialization Phase**: In this phase, the key and the initialization vector (IV) are loaded into the algorithm's states. The states are updated with several clock pulses, and this update process occurs without producing any output. The goal of this update is to combine the key with the IV in such a way that a change in the IV leads to a change in the pseudo-random bit sequence produced by the generator. After the initialization and preparation of the internal states, the algorithm is ready for the next phase, which is the generation of the pseudo-random bit sequence.

2. **Encryption/Decryption Phase**: In this phase, the pseudo-random bit sequence is generated by continuously updating the internal states of the algorithm, and this sequence is used for encryption or decryption by XORing it with the message or ciphertext.

3. **New Session Start**: After several exchanges between the sender and receiver, a new session is initiated with the release of a new initialization vector, although the initial key remains unchanged.

The core of a stream cipher is the pseudo-random bit sequence generator. The most important requirement for this part is that the generated sequence should depend more on the key and less on the initialization vector. To construct a pseudo-random bit sequence generator, suitable structures and mathematical operations, such as Linear Feedback Shift Registers (LFSR) and non-linear feedback mechanisms, can be used. These structures are well-suited for hardware implementation and have good statistical properties.

### 2.4.3 Comparison

Each encryption method, whether block cipher or stream cipher, has its own set of advantages and disadvantages, which can be applied depending on their use case and environment. Table 2.2 summarizes these pros and cons.

Table 2.2: Comparison of Features Between Block and Stream Ciphers

| Feature | Stream Cipher | Block Cipher |
|---|---|---|
| **Message Length** | Variable | Constant |
| **memory** | The algorithm has internal memory. Encryption does not depend only on the message and key, but also on previous changes made to the encoder, which are stored in memory and affect the new output. | Cryptography is an in-place operation, and the result is not stored anywhere. The output does not depend on previous states or data. |
| **Design Model** | Finite State Machine | Confusion+Diffusion |
| **Mathematical equvalence** | Psudeorandom Generator | Psudeorandom Permutation |
| **Speed** | In hardware applications: mostly fast In software applications: faster | In hardware applications: low speed |
| **Core** | Encryption=Decryption | Encryption+Decryption |
| **Implementation Complexity** | In hardware: They have low implementation complexity. In software: Their correct implementation is difficult in both hardware and software, and they are vulnerable to weaknesses based on the application. Improper implementation causes severe weaknesses in the application. | In hardware: They have high implementation complexity. |
| **Buffering** | No need for buffering | Need for buffering |
| **Noise and Error Propagation** | Propagate noise | They do not propagate errors. They are suitable for noisy channels. |
| **Supporting Security Services** | Confidentiality Yes Authentication No Integrity No Non-repudiation No | Confidentiality Yes Authentication Yes Integrity Yes Non-repudiation No |
| **Cost** | Low cost Most algorithms are designed for a specific application. | Costy |
| **Application** | ꞁSecure Wireless Communications Resource-Limited Devices | They become important where communication security is critical. |

From this table, it can be inferred that block ciphers tend to be slower and have high hardware implementation costs, but they are less vulnerable to environmental factors such as noise and connection interruptions. On the other hand, stream ciphers, although faster and cheaper in terms of implementation, do not support data integrity, making them more vulnerable to incorrect implementations and security attacks.

Stream ciphers are more suitable for applications where speed is important, such as in real-time communications or on devices with limited computational resources, while block ciphers are more appropriate in environments requiring high security, especially where data integrity and authentication are critical.

## 2.5 Existing Lightweight Symmentric Encryption Algorithms

Reviewing previous designs of lightweight symmetric encryption algorithms helps to identify the strengths and weaknesses of each algorithm. It is essential to remember that each encryption algorithm is designed with specific intentions and for a particular use case. The following examines the AEAD (Authenticated Encryption with Associated Data) algorithm and several lightweight cipher algorithms along with their features.

### 2.5.1 Present Cipher

The key features of the PRESENT cipher, as cited from various sources, are as follows:

- It is a very lightweight block cipher designed by Axel Yorke-Pashman in 2007 and was chosen as a lightweight encryption standard in 2012.

- The block size is 64 bits, and the key sizes are either 80 or 128 bits.

- The algorithm uses the SPN (Substitution-Permutation Network) structure with 31 rounds.

- It was designed with a focus on hardware implementation.

- Security analysis indicates that attacks on the algorithm structure are effective up to a maximum of 26 rounds, and no effective attacks have been reported on its key generation function.

- Top Level Algorithm Descripton of algorithm is shown in Figure 2.6.

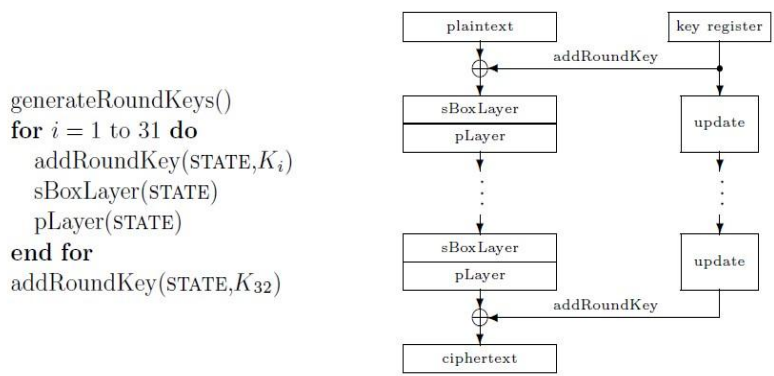

**Fig 2.6: Top Level Algorithm Descripton of Present Cipher**

## 2.5.2 Boron Cipher

The key features of the Boron cipher, as cited by various sources, are as follows:

- It is a block cipher introduced in 2017 by Abahitaj at the University of Pune, India.

- The design of the Bouron cipher is based on the design principles of the PRESENT cipher, and its round structure and key generator are similar to those of the PRESENT algorithm.

- Security analysis of this algorithm has been conducted by the author himself.

- Top Level Algorithm Descripton of the algorithm is shown in Figure 2.7.

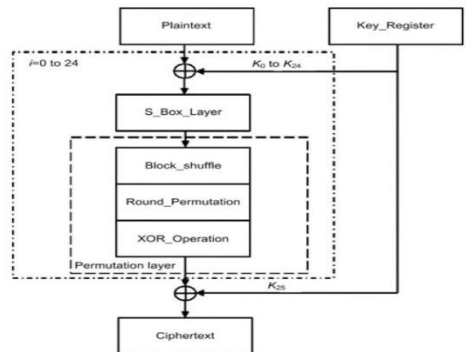

**Fig 2.7: Top Level Algorithm Descripton of Boron Cipher**

## 2.5.3 Hummingbrid Cipher

The key features of the Hummingburd cipher, as cited from various sources [18][19][20], are as follows:

- It is the only ultra-lightweight combined cipher designed by Zhen Zhen Fan in 2010.

- The cipher has a key length of 256 bits and 80-bit internal states. Upon analysis, the algorithm was found to be vulnerable to cryptanalysis. Additionally, the small size of its internal states makes it more vulnerable to stream cipher attacks.

• The improved version of this algorithm, called Hummingburd 2, has a block size of 16 bits, a key length of 128 bits, and a 128-bit state set, which is initialized by a 64-bit initialization vector (IV).

• The design logic of this algorithm is based on the Enigma machine. The encryption process can be viewed similarly to a rotating machine, where the block cipher components perform substitution on the 16-bit word.

• The Hummingburd2 algorithm is proprietary.

• Top Level Algorithm Descripton of the algorithm's structure is shown in Figure 2.8.

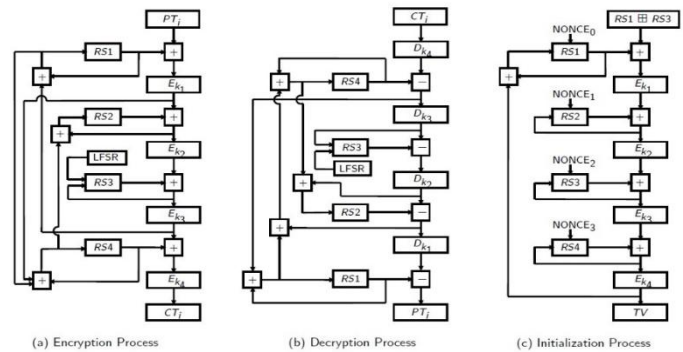

**Fig 2.8: Top Level Algorithm Descripton of Hummingbird Encryption and Decryption**

## 2.5.4 Comparison

The comparison of the previously designed ciphers—Present, Boron, and Hummingbird—is shown in Table 2.3:

**Table 2.3**: **Comparison of Several Common Lightweight Ciphers**

| Cipher | Block Size | Key Length | Structure | Year |
|---|---|---|---|---|
| Present | 64 | 128-bit | Block | 2007 |
| Boron | 64 | 128-bit | Block | 2018 |
| Humingbird | 16 | 256-bit | Stream-Block | 2014 |

## Summary and Conclusion

In this chapter, we reviewed various cryptographic systems, compared their structures, and examined different fundamental encryption methods, including their advantages and disadvantages. This provided us with a general overview of these systems. Studying each cipher algorithm and understanding its key features and applications will help us in designing our own new algorithm. In the next chapter, we will apply the principles and features gained from the reviewed algorithms to propose a new algorithm based on the results obtained.

# Chapter 3: Design and Implementation of the Proposed Algorithm

**Chapter Objectives:**

1. Conceptualization of the SEPAR Algorithm Design

2. Implementation of the SEPAR Algorithm

3. Evaluation of the Implementation Results

# Introduction

In this chapter, we present a newly designed lightweight algorithm named "SEPAR." The primary objective of its design is its application in devices used on the Internet of Things (IoT) environment, including ultra-lightweight devices.

## 3.1 Design Strategy

The main challenge in designing a lightweight cipher is achieving a balance between three parameters: security, cost, and performance. The ideal goal in the design of lightweight algorithms is to create and generalize innovative and novel algorithms that are implemented in a small space, with high speed, low power consumption, and, at the same time, security. Figure 3.1 illustrates the perspective of achieving balance in the design of a lightweight cryptographic algorithm [21].

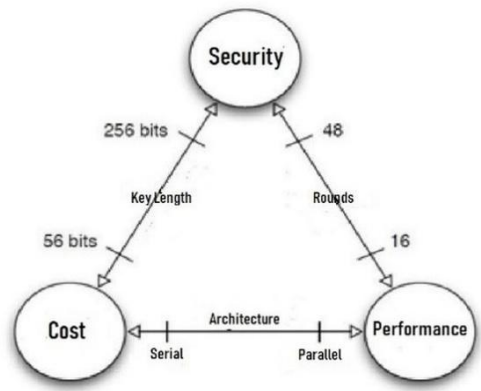

Fig 3.1: Design Trade-offs for lightweight cryptographic algorithms[23]

In the design of a lightweight cryptographic algorithm, we must always maintain a balance between the three parameters mentioned in order to provide an optimized design. An example of the thought process for achieving this balance can be stated as follows: as the key size increases, the provided security increases, but this leads to a decrease in performance due to the rise in key processing rounds, resulting in higher costs in the design.

It should be noted that lightweight cryptographic algorithms use mathematical operators that have a low computational burden. The use of finite field computations in the design of lightweight algorithms is not necessarily due to the high computational power required. Additionally, four-bit S-box should always be used instead of eight-bit boxes to optimize memory usage.

## 3.2 SEPAR Cryptographic System

SEPAR cipher is designed by combining block ciphers and stream ciphers. Its input length is 16 bits, key length is 256 bits, and its internal memory length is 144 bits. The appropriate choice of key length and internal state size not only suits most resource-constrained devices but also provides sufficient security. Table 3.1 contains the common notations used in this cryptographic system.

Table 3.1: Notations used in the description of SEPAR cipher

| Notation | Description |
|---|---|
| KEY | the 256-bit master key |
| $K_i$ | the 32-bit segment key which drives by dividing the master key to 8 equal segments and used in $Enc\_block_i$, i = 1,2, … ,8 |
| $k_i$ | the 16-bit subkey used in each Enc_block, i =1,2, … ,6 |
| State_i | the i-th state which i is number , i = 1,2, … ,9 |
| PT | the 16-bit plaintext |
| CT | the 16-bit ciphertext |
| XOR | Exclusive-or (XOR) operator |
| ⟫n | Right circular shift operator, which rotates all bits of its input to the right by n bits |
| ⟪n | Left circular shift operator, which rotates all bits of its input to the left by n bits |
| ⊞ | module $2^{16}$ addition operator |
| ⊟ | module $2^{16}$ subtraction operator |

## 3.2.1 Initialization Process

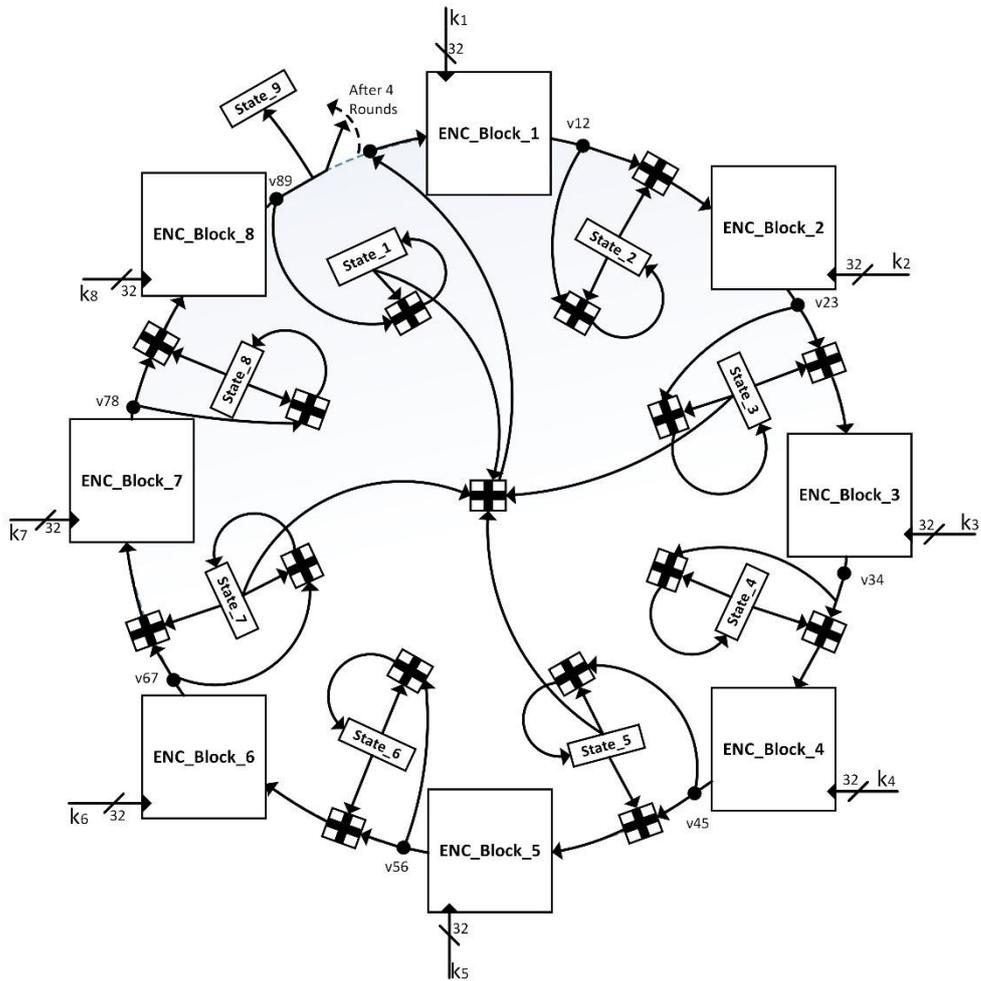

**Fig 3.2: SEPAR initialization algorithm**

The Overall structure of the initialization algorithm can be seen in Figure 3.2. Before starting the encryption process, the initialization algorithm is executed. During the execution of the initialization algorithm, eight random 16-bit values are placed into the eight internal states of the initialization algorithm and are processed for four consecutive rounds. At the end of the execution, the eight current states of the initialization algorithm are copied into the eight internal states of the encryption algorithm. Additionally, the final value generated at the end of the fourth round, after modifying its 7th bit to 1, is stored in state number 9 of the encryption algorithm. The pseudocode of the initialization algorithm process is shown in Pseudocode (3.1).

**Algorithm 2: SEPAR Initialization Algorithm**

**Input**: Eight 16-bit random integers known as a nonce
**Output**: Initialized eight states State_i (i=1,2, 3,…,8) and LFSR

1: $state_1 = NONCE_1$
2: $state_2 = NONCE_2$
3: $state_3 = NONCE_3$
4: $state_4 = NONCE_4$
5: $state_5 = NONCE_5$
6: $state_6 = NONCE_6$
7: $state_7 = NONCE_7$
8: $state_8 = NONCE_8$
9: **for** t = 0 to 3 **do**
10: $\quad V12_t = Enc\_Block_{k_1}(((state1_t \boxplus state3_t) \boxplus state5_t) \boxplus state7_t)$
11: $\quad V23_t = Enc\_Block_{k_2}(V12_t \boxplus state2_t)$
12: $\quad V34_t = Enc\_Block_{k_3}(V23_t \boxplus state3_t)$
13: $\quad V45_t = Enc\_Block_{k_4}(V34_t \boxplus state4_t)$
14: $\quad V56_t = Enc\_Block_{k_5}(V45_t \boxplus state5_t)$
15: $\quad V67_t = Enc\_Block_{k_6}(V56_t \boxplus state6_t)$
16: $\quad V78_t = Enc\_Block_{k_7}(V67_t \boxplus state7_t)$
17: $\quad Out_t = Enc\_Block_{k_8}(V78_t \boxplus state8_t)$
18: $\quad state1_{t+1} = state1_t \boxplus Out_t$
19: $\quad state2_{t+1} = state2_t \boxplus V12_t$
20: $\quad state3_{t+1} = state3_t \boxplus V23_t$
21: $\quad state4_{t+1} = state4_t \boxplus V34_t$
22: $\quad state5_{t+1} = state5_t \boxplus V45_t$
23: $\quad state6_{t+1} = state6_t \boxplus V56_t$
24: $\quad state7_{t+1} = state7_t \boxplus V67_t$
25: $\quad state8_{t+1} = state8_t \boxplus V78_t$
26: **end for**
27: $LFSR = Out_3 \mid 0x100$
28: **return** $statei_7 \ (i = 1.\,...\,.8)$ and LFSR

## 3.2.2 Encryption Process

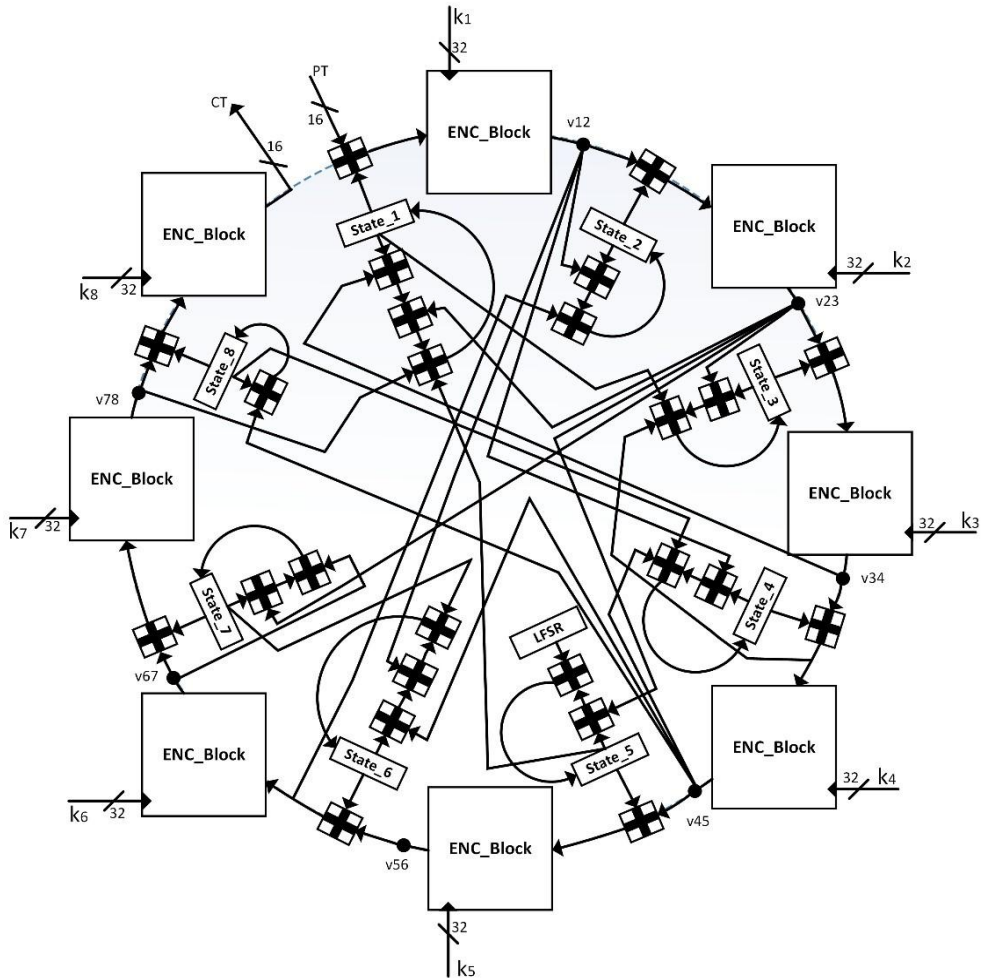

**Fig 3.3: SEPAR Encryption Process**

The overall structure of the encryption algorithm is shown in Figure 3.3. After completing the preparation process, the encryption process is carried out by this algorithm. Pseudocode (3.2) clarifies the workflow of this algorithm.

**Pseudocode (3.2): SEPAR encryption algorithm**

**Input**: A 16-bit plaintext (PT), eight 16-bit states (State_i) and LFSR
**Output**: A 16-bit ciphertext (CT)

1: $V12_t = \text{Enc\_Block}_{k_1}(PT \boxplus state1_t)$
2: $V23_t = \text{Enc\_Block}_{k_2}(V12_t \boxplus state2_t)$
3: $V34_t = \text{Enc\_Block}_{k_3}(V23_t \boxplus state3_t)$
4: $V45_t = \text{Enc\_Block}_{k_4}(V34_t \boxplus state4_t)$
5: $V56_t = \text{Enc\_Block}_{k_5}(V45_t \boxplus state5_t)$
6: $V67_t = \text{Enc\_Block}_{k_6}(V56_t \boxplus state6_t)$
7: $V78_t = \text{Enc\_Block}_{k_7}(V67_t \boxplus state7_t)$
8: $CT_i = \text{Enc\_Block}_{k_8}(V78_t \boxplus state8_t)$
9: $LFSR_{t+1} \leftarrow LFSR_t$
10: $state2_{t+1} = V12_t \boxplus V56_t \boxplus state6_t$
11: $state3_{t+1} = V23_t \boxplus state4_{t+1} \boxplus state1_t$
12: $state4_{t+1} = V12_t \boxplus V45_t \boxplus state8_t$
13: $state5_{t+1} = V23_t \boxplus LFSR_{t+1}$
14: $state6_{t+1} = V12_t \boxplus V45_t \boxplus state7_t$
15: $state7_{t+1} = V23_t \boxplus V67_t$
16: $state8_{t+1} = V45_t$
17: $state1_{t+1} = V34_t \boxplus V23_t \boxplus V78_t \boxplus state5_t$
18: **return** CT

### 3.2.3 Decryption Process

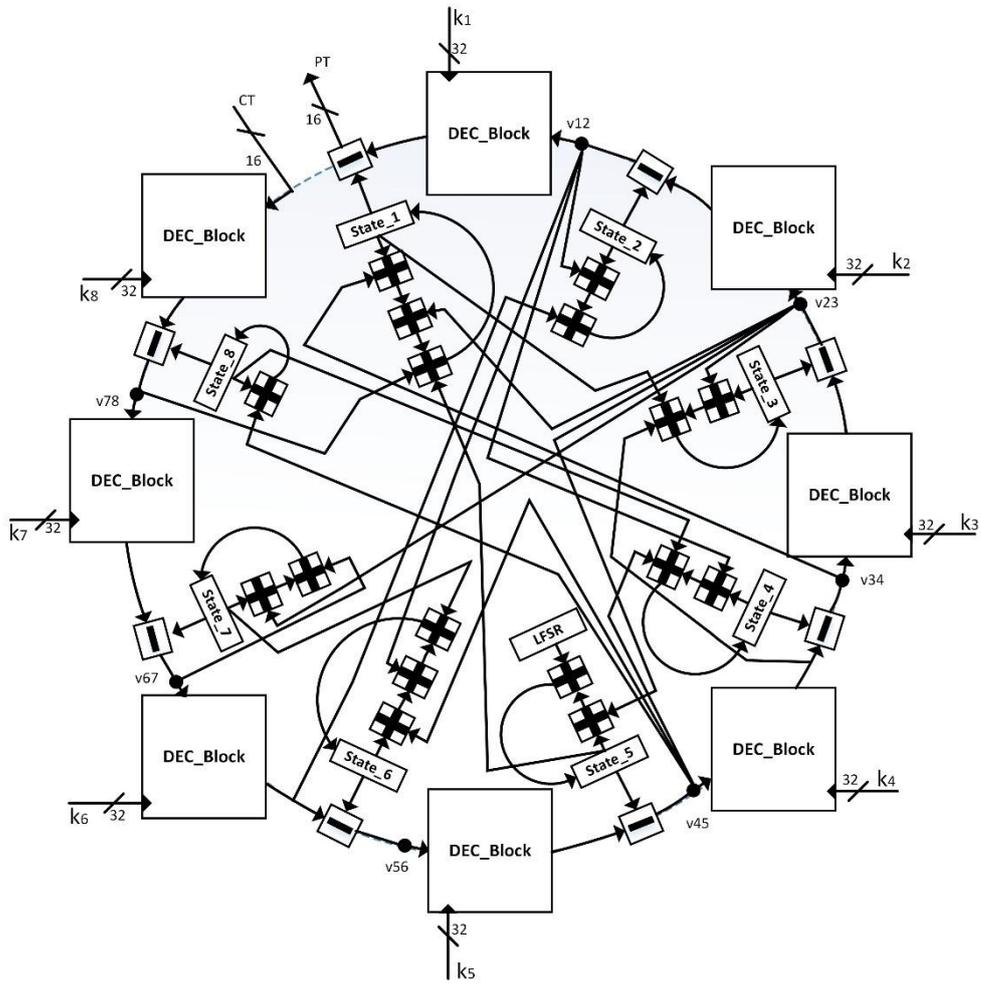

**Fig 3.4: SEPAR Decryption process**

The overall structure of the decryption algorithm is shown in Figure 3.4. The decryption process follows the same pattern as the encryption process but is executed in reverse. The details of the algorithm are provided in pseudocode (3.3).

<div style="text-align:center">Pseudocode (3.3): **SEPAR decryption algorithm**</div>

**Input**: A 16-bit ciphertext (CT) and eight 16-bit states(State_i)
**Output**: A 16-bit plaintext(PT)

1: $V78_t = \text{DEC\_Block}_{k_8}(CT) \boxminus state8_t$
2: $V67_t = \text{DEC\_Block}_{k_7}(V78_t) \boxminus state7_t$
3: $V56_t = \text{DEC\_Block}_{k_6}(V67_t) \boxminus state6_t$
4: $V45_t = \text{DEC\_Block}_{k_5}(V56_t) \boxminus state5_t$
5: $V34_t = \text{DEC\_Block}_{k_4}(V45_t) \boxminus state4_t$
6: $V23_t = \text{DEC\_Block}_{k_3}(V34_t) \boxminus state3_t$
7: $V12_t = \text{DEC\_Block}_{k_2}(V23_t) \boxminus state2_t$
8: $PT_i = \text{DEC\_Block}_{k_1}(V12_t) \boxminus state1_t$
9: $LFSR_{t+1} \leftarrow LFSR_t$
10: $state2_{t+1} = V12_t \boxplus V56_t \boxplus state6_t$
11: $state3_{t+1} = V23_t \boxplus state4_{t+1} \boxplus state1_t$
12: $state4_{t+1} = V12_t \boxplus V45_t \boxplus state8_t$
13: $state5_{t+1} = V23_t \boxplus LFSR_{t+1}$
14: $state6_{t+1} = V12_t \boxplus V45_t \boxplus state7_t$
15: $state7_{t+1} = V23_t \boxplus V67_t$
16: $state8_{t+1} = V45_t$
17: $state1_{t+1} = V34_t \boxplus V23_t \boxplus V78_t \boxplus state5_t$
18: **return** PT

## 3.2.4 Key Generation Algorithm

SEPAR algorithm can be implemented without a fixed key generator, and the subkeys can be generated on the fly; However for enhanced security, a very lightweight key generator is proposed, which is designed with time and memory optimization in mind, while also ensuring resistance to key generation attacks. The pseudocode for generating subkeys is presented in pseudocode (3.4).

<div style="text-align:center">pseudocode (3.4): **Key Generator Algorithm**</div>

**Input:** A 256-bit master key(KEY) and Number of Enc_block(n)
**Output:** Six 16-bit Subkeys(**k**_i_)

1: $KEY = K_1 \| K_2 \| K_3 \| K_4 \| K_5 \| K_6 \| K_7 \| K_8$
2: $K_j = k_1 \| k_2 \quad . j = 1.2.3.4.5.6.7.8$
3: **k₁** $= k_1$
4: **k₂** $= k_2$
5: **k₃** $= SBOX((k_1 \lll 6)[7.8.9.10]) \oplus n + 2$
6: **k₄** $= SBOX((k_2 \lll 10)[7.8.9.10]) \oplus n + 3$
7: **k₅** $= k_1 \oplus k_2$
8: **k₆** $=$ **k₃** $\oplus$ **k₄**
9: **return k**_i_ $\quad . i = 1.2.3.4.5.6$

## 3.2.5 16-bit Block Cipher

In the encryption process, SEPAR algorithm uses eight 16-bit blocks, referred to as **Enc-blocks**. Each **Enc-block** employs a block cipher function called **b16** in its structure. The internal structure of **b16** is an SP network, as shown in Figure 3.5. The substitution layer is implemented using four S-box (S-boxes).

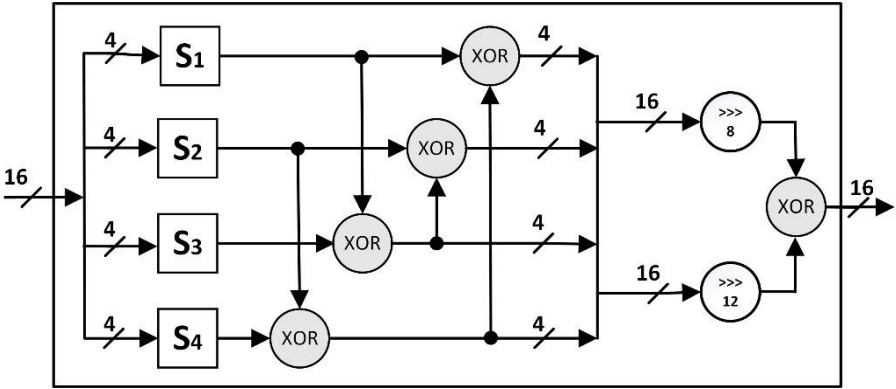

**Fig 3.5: b16 block cipher**

Each Enc-block consists of a consecutive sequence of four b16 encoders and a substitution box layer. The hexadecimal representation of the four S-box used in b16 is provided in Table 3-2.

**Table 3-2: Selected Four S-boxes in Hexadecimal Notation**

| x | 0 | 1 | 2 | 3 | 4 | 5 | 6 | 7 | 8 | 9 | A | B | C | D | E | F |
|---|---|---|---|---|---|---|---|---|---|---|---|---|---|---|---|---|
| $S_1(x)$ | 1 | F | B | 2 | 0 | 3 | 5 | 8 | 6 | 9 | C | 7 | D | A | E | 4 |
| $S_2(x)$ | 6 | A | F | 4 | E | D | 9 | 2 | 1 | 7 | C | B | 0 | 3 | 5 | 8 |
| $S_3(x)$ | C | 2 | 6 | 1 | 0 | 3 | 5 | 8 | 7 | 9 | B | E | A | D | F | 4 |
| $S_4(x)$ | D | B | 2 | 7 | 0 | 3 | 5 | 8 | 6 | C | F | 1 | A | 4 | 9 | E |

The choice of the four S-box is based on the results obtained from the research [22], which, through extensive analysis of desirable 4-bit S-box, identified a set of S-box called "golden S-box." The security features present in these boxes ensure that encryption algorithms remain resilient to most significant attacks, particularly linear and differential attacks. In subsection 3-2-5-1, we further explore these criteria. The permutation layer used in **b16** is implemented with simple bitwise addition and rotation operators, providing adequate diffusion properties. The key mixing stage in this structure is straightforward, achieved with a bitwise addition operation. The four extracted 16-bit subkeys (**k_1, k_2, k_3, k_4**) from the 32-bit key section are bitwise added to each **b16**. Additionally, the two 16-bit subkeys (**k_5, k_6**) are bitwise added to their corresponding substitution layer within the **Enc-block**.

Figure 3.6 illustrates the overall structure of the **Enc-block** structure, and the pseudocode for the execution of the **Enc-block** operation is provided in pseudocode (3.5).

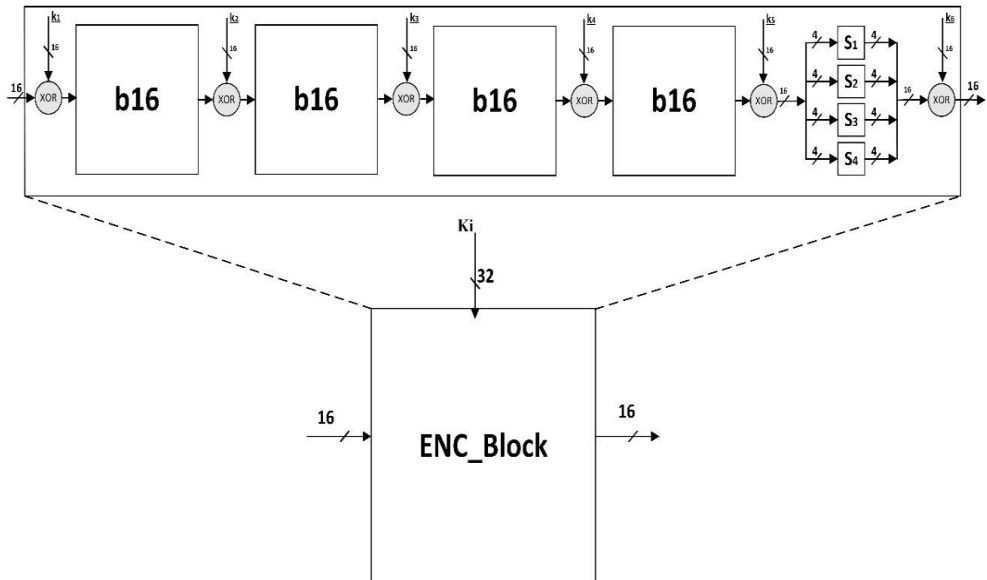

**Fig 3.6: Enc_block Structure**

Pseudocode (3.5): **Enc_block operation process**

1. **Input**: 16-bit input (m) and six 16-bit subkeys (**k_i**)
2. **Output**: A 16-bit data block (m´)
3. 
4. 1: **for** j = 1 to 4 **do**
5. 2:    m ← m ⊕ **k_j**
6. 3:    $A = m_0||m_1||m_2||m_3$   $B = m_4||m_5||m_6||m_7$
7.        $C = m_8||m_9||m_{10}||m_{11}$   $D = m_{12}||m_{13}||m_{14}||m_{15}$
8. 4:    A ← S(A)   B ← S(B)   C ← S(C)   D ← S(D)
9. 5:    A ← A ⊕ C   B ← B ⊕ D   C ← C ⊕ B   D ← D ⊕ A
10. 6:   m ← A||B||C||D
11. 7:   m ← m ⊕ (m ⋘ 8) ⊕ (m ⋘ 12)
12. 8: **end for**
13. 9: m ← m ⊕ **k_5**
14. 10: $A = m_0||m_1||m_2||m_3$   $B = m_4||m_5||m_6||m_7$
15.     $C = m_8||m_9||m_{10}||m_{11}$   $D = m_{12}||m_{13}||m_{14}||m_{15}$
16. 11: m ← S(A)||S(B)||S(C)||S(D)
17. 12: m´ ← m ⊕ **k_6**
18. 13: **return** m´
19.

## 3.2.5.1 Criteria for chosing S-Boxes

A 4-bit to 4-bit substitution box, is a Boolean vector function $S(x) : \mathbb{F}_2^4 \rightarrow \mathbb{F}_2^4$. Suppose $x = (x_3||x_2||x_1||x_0)$ is a 4-bit input to the substitution box $S(x)$. S-box can be represented in its boolean form as shown in (3-1):

$$S(x) = S^{(3)}(x)||S^{(2)}(x)||S^{(1)}(x)||S^{(0)}(x) \qquad (3\text{-}1)$$

Each of the $S^{(i)}(x)$ is a component function of the substitution box S(x). S-box play a crucial role in introducing the nonlinearity feature in a cryptographic algorithm, and their proper selection enhances the algorithm's resistance against most important cryptographic analyses and attacks. In the context of examining and analyzing 4-bit S-box, two fundamental and extensive studies in [23] were conducted, categorizing 4-bit S-box based on their linear and differential properties. Based on the results from the study [23], the categorization of S-box introduced a desirable category that ensures immunity against most major cryptographic attacks. Key features of the selected S-box include the following:

- **Degree of Immunity against Differential Attack**: For a substitution box $S(x)$ with an input size of $|s|$, the degree of immunity against a differential attack is defined as the number of elements of x for which the following relation:

$$S(x + \Delta_i) = S(x) + \Delta_o \qquad (3\text{-}2)$$

- **Maximum Differential Probability** is given by:

$$P = \frac{n}{|s|} \qquad (3\text{-}3)$$

  This is calculated using relation (3-3). For an ideal S-box, the lower the maximum differential probability, the better. In golden S-boxes, this value is $\frac{1}{4}$. This feature is effective for countering differential attacks in the algorithm.

- **Maximum Linear Probability**: One of the most important and fundamental features a substitution box must have is its nonlinearity. The nonlinearity requirement of a substitution box ensures that there is no linear mapping between its inputs and outputs. This feature allows us to define a measure of nonlinearity for the given element. For a substitution box, the linearity property can be evaluated by checking the following relation (3-4):

$$(x)\&(input\ mask) \stackrel{?}{==} S(x)\&(output\ mask) \qquad (3\text{-}4)$$

In equation (3-4), the input mask is calculated as $(x_3 \oplus x_2 \oplus x_1 \oplus x_0)$ and the output mask is calculated as $(y_3 \oplus y_2 \oplus y_1 \oplus y_0)$, where each of $x_i$ and $y_i$ is a bit number. The number of different expressions for all input masks and output masks is calculated and after subtracting the value of eight, it is collected in a table called the linear approximation table. The maximum linear probability for this substitution box is calculated similarly to the calculation of the maximum differential probability, from equation (3-4), except that n is the largest number of expressions whose input mask is equal to their output. For an optimal substitution box, the lower the maximum linear probability, the better. In golden S-box, this value is equal to $\frac{1}{4}$. This feature is effective for dealing with linear attacks in the algorithm.

- **Algebraic Degree**: The algebraic degree of a Boolean function is the highest degree of its algebraic normal form components. This feature makes the algorithm resistant to algebraic attacks. For a golden substitution box, this value is 3.

  In the following, we will analyze each of these features. For this purpose, $S_1(x)$ is chosen as an example. Using Online Minimization of Boolean Functions, the components of the selected substitution box are expressed as Boolean functions in the relation (3-5).

$$S(x) = S^{(3)}(x) || S^{(2)}(x) || S^{(1)}(x) || S^{(0)}(x) \quad (3\text{-}5)$$

$$S^{(0)}(x) = \bar{x}_3\bar{x}_2\bar{x}_1 x_0 + \bar{x}_3\bar{x}_2 x_1 \bar{x}_0 + \bar{x}_3 x_2 \bar{x}_1 x_0 + \bar{x}_3 x_2 x_1 x_0$$
$$+ x_3 \bar{x}_2 \bar{x}_1 \bar{x}_0 + x_3 \bar{x}_2 \bar{x}_1 x_0 + x_3 x_2 \bar{x}_1 x_0$$
$$+ x_3 x_2 x_1 \bar{x}_0$$

$$S^{(1)}(x) = \bar{x}_3 \bar{x}_2 \bar{x}_1 x_0 + x_3 \bar{x}_2 x_1 \bar{x}_0 + \bar{x}_3 x_2 x_1 \bar{x}_0 + \bar{x}_3 x_2 x_1 x_0$$
$$+ x_3 \bar{x}_2 x_1 \bar{x}_0 + x_3 x_2 \bar{x}_1 \bar{x}_0 + x_3 x_2 \bar{x}_1 x_0$$
$$+ x_3 x_2 x_1 x_0$$

$$S^{(2)}(x) = \bar{x}_3 \bar{x}_2 \bar{x}_1 x_0 + \bar{x}_3 x_2 \bar{x}_1 x_0 + \bar{x}_3 x_2 x_1 \bar{x}_0 + x_3 \bar{x}_2 \bar{x}_1 \bar{x}_0$$
$$+ x_3 \bar{x}_2 x_1 x_0 + x_3 x_2 \bar{x}_1 x_0 + x_3 x_2 x_1 \bar{x}_0$$
$$+ x_3 x_2 x_1 x_0$$

$$S^{(3)}(x) = \bar{x}_3 \bar{x}_2 \bar{x}_1 x_0 + \bar{x}_3 \bar{x}_2 x_1 \bar{x}_0 + \bar{x}_3 \bar{x}_2 x_1 x_0 + \bar{x}_3 x_2 \bar{x}_1 \bar{x}_0$$
$$+ \bar{x}_3 x_2 \bar{x}_1 x_0 + x_3 \bar{x}_2 \bar{x}_1 \bar{x}_0 + x_3 x_2 \bar{x}_1 \bar{x}_0$$
$$+ x_3 x_2 x_1 x_0$$

In this representation $\bar{x}_i$, denotes the inverse of the bit $x_i$, and the operator represents the logical XOR (exclusive OR) operation. Based on this representation, it is clear that there are 32 terms, and the number of terms for each component of the S-box is equal to 8. The differential distribution table for the S-box used is shown in Figure 3-7.

$$\begin{pmatrix} 16 & 0 & 0 & 0 & 0 & 0 & 0 & 0 & 0 & 0 & 0 & 0 & 0 & 0 & 0 & 0 \\ 0 & 0 & 0 & 2 & 0 & 4 & 2 & 0 & 0 & 0 & 0 & 2 & 2 & 2 & 0 & 2 \\ 0 & 0 & 0 & 0 & 0 & 2 & 0 & 2 & 0 & 2 & 4 & 2 & 0 & 0 & 0 & 4 \\ 0 & 0 & 2 & 2 & 4 & 0 & 0 & 0 & 2 & 2 & 0 & 0 & 0 & 0 & 0 & 4 \\ 0 & 0 & 0 & 0 & 0 & 0 & 4 & 4 & 0 & 0 & 2 & 2 & 2 & 2 & 0 & 0 \\ 0 & 2 & 0 & 2 & 0 & 0 & 0 & 2 & 2 & 2 & 2 & 2 & 0 & 2 & 0 \\ 0 & 2 & 0 & 2 & 0 & 0 & 2 & 2 & 4 & 0 & 0 & 2 & 0 & 0 & 0 & 2 \\ 0 & 0 & 2 & 4 & 4 & 2 & 0 & 0 & 2 & 0 & 0 & 0 & 0 & 2 & 0 \\ 0 & 0 & 0 & 2 & 0 & 0 & 2 & 0 & 0 & 0 & 2 & 4 & 4 & 0 \\ 0 & 2 & 2 & 2 & 0 & 2 & 0 & 2 & 0 & 2 & 2 & 0 & 0 & 0 & 2 \\ 0 & 0 & 2 & 0 & 2 & 0 & 2 & 2 & 0 & 4 & 0 & 2 & 2 & 0 & 0 & 0 \\ 0 & 2 & 2 & 0 & 2 & 2 & 0 & 0 & 2 & 2 & 0 & 2 & 2 & 0 & 0 \\ 0 & 2 & 0 & 0 & 2 & 0 & 2 & 2 & 4 & 0 & 2 & 0 & 0 & 0 & 2 & 0 \\ 0 & 2 & 2 & 0 & 2 & 4 & 0 & 2 & 0 & 0 & 0 & 0 & 0 & 4 & 0 & 0 \\ 0 & 4 & 2 & 0 & 0 & 2 & 0 & 0 & 0 & 0 & 2 & 0 & 2 & 2 & 2 \\ 0 & 0 & 2 & 0 & 0 & 0 & 2 & 2 & 0 & 2 & 2 & 2 & 0 & 4 & 0 \end{pmatrix}$$

**Figure 37: S-Box differential distribution table**

Based on the differential distribution table, value 4 is the largest number inside this table, and as a result, the maximum differential probability for this S-box will be 0.25.
The linear approximation table for the substitution box used is shown in Figure 3-8.

$$\begin{pmatrix} 8 & 0 & 0 & 0 & 0 & 0 & 0 & 0 & 0 & 0 & 0 & 0 & 0 & 0 & 0 & 0 \\ 0 & 2 & 0 & -2 & 0 & 2 & -4 & 2 & 0 & 2 & 0 & -2 & 0 & 2 & 4 & 2 \\ 0 & -2 & 2 & 0 & 0 & 2 & 2 & -4 & -2 & 0 & 0 & 2 & 2 & 0 & 4 & 2 \\ 0 & 4 & 2 & 2 & 0 & 0 & -2 & 2 & -2 & -2 & 0 & 4 & 2 & -2 & 0 & 0 \\ 0 & 0 & 2 & 2 & 2 & 2 & 0 & 0 & 0 & 4 & 2 & -2 & 2 & -2 & 0 & -4 \\ 0 & -2 & -2 & 0 & 2 & 0 & 0 & 2 & 0 & 2 & -2 & 4 & 2 & 4 & 0 & -2 \\ 0 & -2 & 0 & -2 & -2 & 0 & 2 & 4 & 2 & 0 & 2 & 0 & 4 & -2 & 0 & 2 \\ 0 & 0 & 4 & 0 & -2 & 2 & 2 & 2 & 2 & 2 & -2 & 2 & -4 & 0 & 0 & 0 \\ 0 & 0 & 0 & 4 & 2 & -2 & 2 & 2 & -2 & 2 & -2 & -2 & 0 & 0 & 0 & 4 \\ 0 & 2 & 0 & 2 & -2 & -4 & 2 & 0 & 2 & 0 & 2 & 0 & 0 & 2 & 4 & -2 \\ 0 & 2 & 2 & 0 & -2 & 0 & 0 & -2 & 0 & 2 & 2 & 0 & 2 & 4 & -4 & 2 \\ 0 & 0 & 2 & 2 & 2 & 2 & 0 & 0 & 4 & -4 & -2 & -2 & 2 & 2 & 0 & 0 \\ 0 & 0 & -2 & 2 & 0 & 4 & 2 & 2 & -2 & -2 & 4 & 0 & -2 & 2 & 0 & 0 \\ 0 & -2 & 2 & 0 & 4 & -2 & -2 & 0 & 2 & 0 & 4 & 2 & -2 & 0 & 0 & 2 \\ 0 & 2 & -4 & 2 & 0 & 2 & 0 & -2 & 4 & 2 & 0 & 2 & 0 & -2 & 0 & 2 \\ 0 & 4 & 0 & -4 & 4 & 0 & 4 & 0 & 0 & 0 & 0 & 0 & 0 & 0 & 0 & 0 \end{pmatrix}$$

**Figure 3.8: Linear approximation table for the substitution box**

From this table, we can conclude that the maximum linear probability for this substitution box will also be 0.25. Additionally, the algebraic degree of all components of this substitution box in its Boolean expression is at least 3, which ensures that the algorithm's representation in terms of algebraic equations is sufficiently complex. Based on these results, we can say that the chosen substitution box provides the necessary security for the SEPAR algorithm.

### 3.2.6 Criteria for Choosing LFSR

The purpose of using a Linear Feedback Shift Register (LFSR) is to introduce greater complexity into the algorithm's structure while minimizing hardware costs. The LFSR is a bit sequence generator that, when an optimal characteristic polynomial is chosen, produces an output sequence with a maximum period and appropriate statistical properties. The characteristic polynomial for the LFSR used is given by:

$$f(x) = x^{16} + x^{15} + x^{12} + x^{10} + x^7 + x^3 + 1 \qquad (3\text{-}6)$$

This polynomial is an irreducible polynomial with a maximum period of $2^{16} - 1$. In Figure 3-9, a sample output sequence of 800 bits from this Linear Feedback Shift Register is shown.

**Figure 3-9:** a sample output sequence of 800 bits

### 3.2.7 Logic of Enc-Block Design

This cipher is designed by combining block and stream ciphers to address the security weaknesses of stream ciphers while leveraging the security strengths of block ciphers and the high speed of stream ciphers. Its 16-bit block size is chosen due to its compatibility with word-based machine architecture, offering high encryption speed. At the same time, it allows the decryption process to be implemented without the need to calculate the inverse of the b16 function, making decryption feasible in specific implementation models.

### 3.2.8 Security Analysis

In this section, the security analysis of the "Spear" algorithm is provided. The purpose of these analyses is to demonstrate the algorithm's resistance to various well-known cryptographic attacks and analyses. The hybrid structure of this algorithm necessitates the examination of both block cipher and stream cipher attacks and analyses. The open-source mathematical software SageMath (version 8.2) is used for these analyses.

1. **Birthday Attack:** In the birthday attack, the attacker's goal is to find two different sets of states (states 1 to 8 along with the linear feedback shift register) initialized by two distinct initial vectors using the same fixed key. The countermeasure to this attack is to create a one-to-one mapping between the setup algorithm's states and the encryption algorithm's states. In the design of the "Spear" cipher system, this principle is adhered to, ensuring that a one-to-one mapping between these states exists. Additionally, the 144-bit internal state size of the encryption algorithm prevents the birthday paradox from being exploited, making it computationally difficult to find collisions with a probability higher than 50%. This is because testing at least 72 bits of the main key is sufficiently complex from a computational perspective.

2. **Differential Analysis:**

$$D_F(a.b) = |\{F(x) + F(x+a) = b. x \in \mathbb{F}_2^m\}| \qquad (3\text{-}7)$$

In the proposed algorithm, the possible differences for the b16 function are computed for different inputs and shown in Table 3-3. Based on these results, the relationship in Equation (3-7) can be derived.

Table 5: Differential property of the 16-bit block cipher

| # of b16 in Enc_block | $max_{a \neq 0.b} D_{b16}(a.b)$ |
|---|---|
| 0 | 16370 |
| 1 | 1016 |
| 2 | 84 |
| 3 | 22 |
| 4 | 22 |

The analysis conducted from Table 3.3 leads to the conclusion of relation (3.8):

$$max_{a.b \in F\mathbb{F}_2^{16}.a \neq 0}\{D_{b16_i}(a.b)\} \leq 20 \qquad (3\text{-}8)$$

Based on relation (3.8) and Table 3.3, it can be concluded that four consecutive rounds of the b16 cryptographic function provide sufficient indistinguishability to the block cipher "Enc-Block" and ensure adequate security against differential analysis. To further examine the algorithm's resistance to differential attacks, high-probability differential characteristics in consecutive iterations of the b16 cipher function have been evaluated. Table 3.4 displays the number of high-probability differential characteristics for consecutive rounds of the b16 cipher. As shown in this table, in the fifth round of Enc-block, only two high-probability differential characteristics exist, indicating the algorithm's security against this attack.

Table 6: Higher differential characteristics of Enc_block

| # of b16 in Enc_block | # of High probability Differential characteristics | Probability Value |
|---|---|---|
| 1 | 28 | $\frac{1}{4}$ |
| 2 | 9 | $\frac{1}{64}$ |
| 3 | 7 | $\frac{1}{4096}$ |
| 4 | 7 | $\frac{1}{52428}$ |

Two high-probability differential characteristics for five rounds are shown in relation (3.9).

$[0000\ 0011\ 0000\ 0000] \rightarrow [0000\ 0101\ 0000\ 0000]$  (3-9)
$[0000\ 0111\ 0000\ 0000] \rightarrow [0000\ 1101\ 0000\ 0000]$

In Figure 3.10, the path of these high-probability differential characteristics, along with the active S-box, is shown.

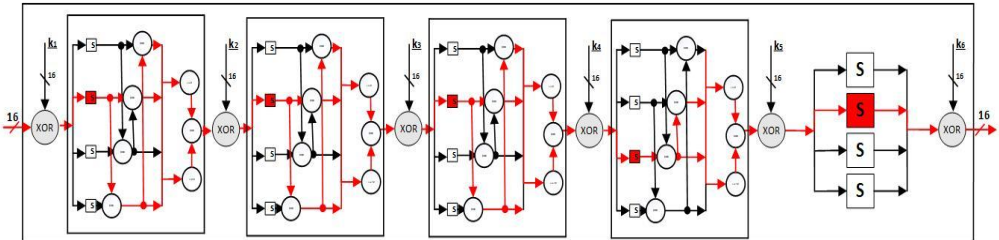

Fig 3. 10: Differential characteristic path

The maximum differential probability of the S-box used is equal to $2^{-2}$. Given the presence of a maximum of 5 active S-box in the enc-block, the differential probability for this Enc-block is $2^{-10}$. The presence of eight consecutive enc-blocks in the encryption algorithm reduces the differential probability to $2^{-80}$. Based on the complexity of the differential attack, which is derived from the relation $N_D = \frac{C}{P_D}$, the number of chosen plaintexts and ciphertexts will be $P_D = 2^{80}$, which is an acceptable value for the security of this cipher against differential attacks. Therefore, the use of eight consecutive enc-blocks in the encryption algorithm's structure ensures that the algorithm is resistant to this attack.

3. **Linear Cyptanalysis**: In the design of the b16 function used in the Shield, high-degree linear S-box are employed. To assess the resistance of the Boolean function used in the enc-block to linear analysis, different cases of the number of active S-box in multiple iterations of B16 within an enc-block can be considered. If we assume that in five consecutive iterations of B16, exactly five active S-box with a linear bias of $2^{-2}$ are present, and in the intermediate B16 expressions, the linear bias is $2^{-3}$ (considering the structure of the enc-block), then the bias of one enc-block would be $2^{-9}$, and for eight consecutive enc-blocks, the bias would be $2^{-12}$. Assuming that each enc-block contains six active S-box instead of five, the bias for each enc-block becomes $2^{-12}$ and for eight consecutive enc-blocks, the bias would be $2^{-96}$. Comparing these two assumptions, it can be concluded that the lower bound for the linear bias of one enc-block is $2^{-72}$ making the complexity of the linear analysis $2^{144}$, requiring $2^{144}$ pairs of ciphertext and known plaintext. This amount of data indicates the algorithm's security against linear attacks.

4. **Algebraic Attack [24]**: The 4-bit to 4-bit S-box can be described with a minimum of 21 equations. In order to completely represent an algorithm algebraically, x = a × 21 quadratic equations in y = a × 8 variables (*a* represents the number of S-boxes used in the encryption and key scheduling algorithm) are used to examine the cipher [2]. In the SEPAR encryption algorithm, each Enc_block uses 18 S-boxes accompanying 2 more S-boxes for generating third and fourth subkeys. For the whole of 16 Enc_block used in the encrypting process (initialization plus encryption phase), $16 \times 18 = 288$ S-boxes were used and $16 \times 2 = 32$ S-boxes used for the third and fourth subkey generating. The number of quadratic equations is given as:

$$a = (288 + 32) \times 21 = 6720 \quad \text{(3-10)}$$

The number of variables is calculated as:

$$b = (288 + 32) \times 8 = 2560 \quad \text{(3-11)}$$

Equation (3, 11) is calculated. Table 3 5 shows a comparison of the number of S-box, binomial equations, and number of variables.

Table 9: computed number of S-boxes, Quadratic equations and variables

| Cipher | # of S-boxes | # of quadratic equations | # of variables |
|---|---|---|---|
| Present-80[3] | 527 | 11067 | **4216** |
| BORON-128[2] | 450 | 9450 | **3600** |
| **SEPAR** | **320** | **6720** | **2560** |
| MIBS-64[20] | 320 | 6720 | **2560** |
| KLEIN-80[21] | 240 | 5040 | **1920** |

Based on this table and considering the 6720 second-degree equations and 2560 variables, along with the use of modulo $2^{16}$ operators in the algorithm's structure, it can be confidently stated that finding efficient linear algebraic expressions for this structure is difficult, and thus the algorithm will be secure against algebraic attacks.

5. **Key-related Attack**: Key-related attacks exploit weaknesses in the key generator algorithm. The key generator used in the Shield is sampled from the present key generator, which has not yet been reported to have any specific attacks. Therefore, the aforementioned attacks are not applicable to this algorithm.
6. **Complementary Properties**: The complementary property for the DES algorithm is such that when a plaintext is encrypted with a key, producing a ciphertext, the encryption of the complement of the plaintext with the complement of the same key will result in the complement of the previously encrypted ciphertext. In the Shield structure, due to the transfer of carry values from one sixteen-bit state to another, the complementary property will not occur.
7. **Avalanche Effect**: The avalanche effect of the Shield algorithm is observed in Table 3.6. This table shows how the ciphertext changes when a single bit of the key or plaintext is modified.

Table 10: Avalanche Effect for SEPAR

|  | Main Test Vector | # of bits changed |
|---|---|---|
| Plaintext | 156F19E18FE6297519A352C45731536A | Base Value |
| Key | E8B9B733DA5D96D702DD3972E95307FD50C512DBF44A233E8D1E9DF5FC7D6371 | Base Value |
| IV | 00000000000000000000000000000000 | Base Value |
| Ciphertext | 41E15D769296494746F638CE27FB07E9 | Base Value |
|  | Test Vector 1 |  |
| Plaintext | <u>0</u>56F19E18FE6297519A352C45731536A | One bit Change |
| Key | E8B9B733DA5D96D702DD3972E95307FD50C51DBF44A233E8D1E9DF5FC7D6371 | Fixed |
| IV | 00000000000000000000000000000000 | Fixed |
| Ciphertext | 3DA5B84A3909A41592229A0600805A74 | # of bits Changed: 70 |
|  | Test Vector 2 |  |
| Plaintext | 156F19E18FE6297519A352C45731536A |  |
| Key | E8B9B733DA5D96D702DD3972E95307FD50C512DBF44A233E8D<u>0</u>E9DF5FC7D6371 | One bit Change |
| IV | 00000000000000000000000000000000 | Fixed |
| Ciphertext | FFC2470821454862CD440E210F2051A7 | # of bits Changed: 69 |
|  | Test Vector 3 |  |
| Plaintext | 156F19E18FE6297519A352C45731536A | Fixed |
| Key | E8B9B733DA5D96D702DD3972E95307FD50C512DBF44A233E8D1E9DF5FC7D6371 | Fixed |
| IV | 0000<u>1</u>000000000000000000000000000 | One bit Change |
| Ciphertext | 0B03EF33C4E57C7C6B1E1F31A133DDDD | # of bits Changed: 65 |

8. **Statistical Tests**: The NIST statistical test suite is used to examine the statistical properties of the ciphertext. This suite includes 16 targeted statistical tests designed for cryptographic algorithm analysis. The results of applying these tests on 100 output samples, each with a length of $10^6$ from this algorithm, demonstrate the appropriate statistical properties of this design [25]. These tests, along with their values and status, are shown in Table 3.7.

Table 11: Results of NIST statistical analysis on SEPAR.

| # | Statistical Test Name | $P-PAVLUE$ | Condition |
|---|---|---|---|
| 1 | Frequency (Monobit) Test | 0.487682 | **PASS** |
| 2 | Frequency Test within a Block Test | 0.20127 | **PASS** |
| 3 | Runs Test | 0.504441 | **PASS** |
| 4 | Test the longest Run of ones in a block Test | 0.622270 | **PASS** |
| 5 | Binary Matrix Rank Test | 0.230424 | **PASS** |
| 6 | Spectral DFT Test | 0.417304 | **PASS** |
| 7 | Non-Overlapping Template Matching Test | 0.513000 | **PASS** |
| 8 | Overlapping Template Matching Test | 0.588514 | **PASS** |
| 9 | Moor's "universal statistical" test | 0.249642 | **PASS** |
| 10 | Linear Complexity Test | 0.790818 | **PASS** |
| 11 | Lempel-Ziv Test | 0.915995 | **PASS** |
| 12 | Serial Test | 0.703427 | **PASS** |
| 13 | Approximation Entropy test | 0.546145 | **PASS** |
| 14 | Cumulative Sums (cusum) Test | 0.697960 | **PASS** |
| 15 | Random Excursions Test | 0.506768 | **PASS** |
| 16 | Random Excursions Variant Test | 0.407024 | **PASS** |

In further examination of the statistical properties, the histograms of SEPAR and AES algorithms for a data sample of $10^6$ are shown in Figures 3.11 and 3.12. The histogram for SEPAR demonstrates a distribution close to uniform for the characters. Given the assurance of a perfectly uniform output for AES and the inability to distinguish between these two charts, it can be concluded that the SEPAR algorithm exhibits uniformity.

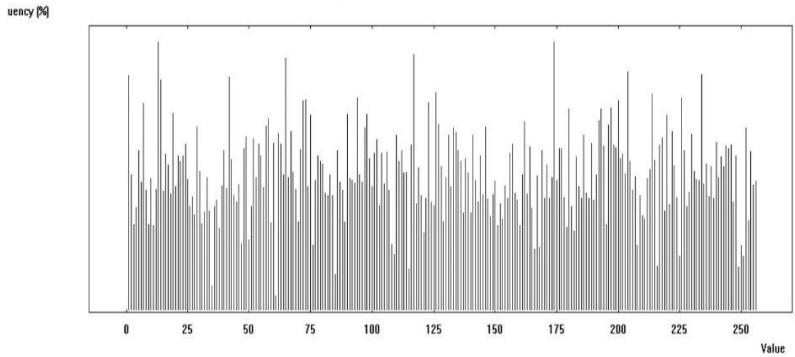
**Figure 3.11: Histogram for SEPAR Algorithm**

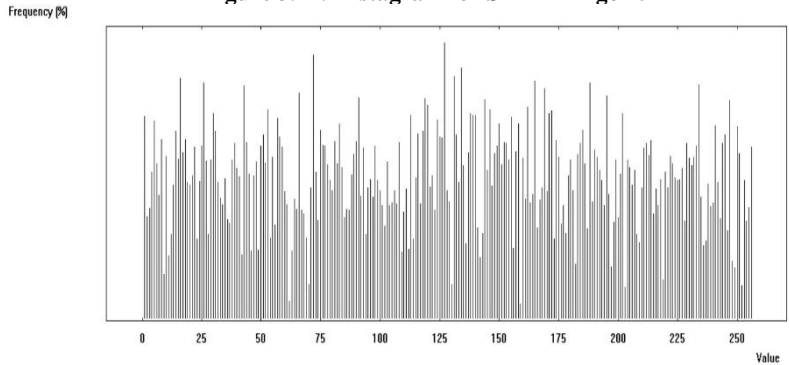
Figure 3.12: Histogram for AES Algorithm

The periodicity feature [14] determines the presence of repetitive sequences of strings longer than one in the ciphertext. In the examination of this feature for the SEPAR algorithm, no periodic repetition of characters was found in various ciphertext samples.

Autocorrelation [13] for a text is an indicator used to measure the similarity between different datasets within the same text. The autocorrelation chart for a SEPAR cipher output sample is shown in Figure 3.13. This chart indicates that no similarity is observed between the sections of the ciphertext.

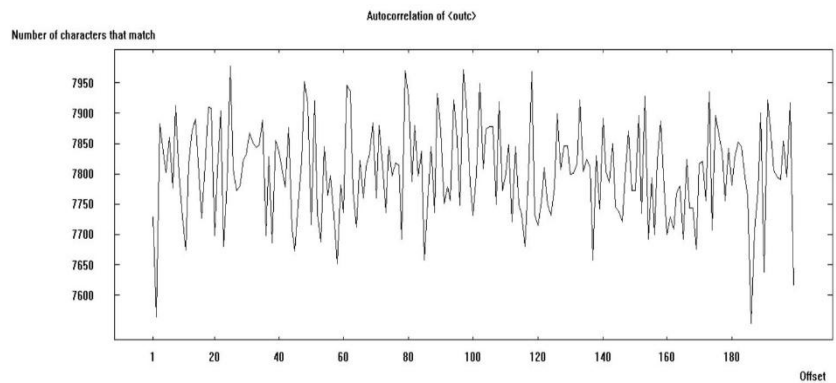
**Fig 3.13: Autocorrelation Diagram**

The entropy feature [13] is a measure used to assess the level of disorder or randomness in a text. For a ciphertext of length $10^6$ that consists of 256 characters, the entropy values are calculated based on Equation (3.12) and are compiled in Table 3.8

$$H(x) = -\sum_{i=1}^{n} p_i log_2^{p_i} \qquad (3\text{-}12)$$

Table 3.8: Entropy comparison

| Cipher Name | Entropy |
|---|---|
| SEPAR | 7.99 |
| AES | 7.99 |
| MARS | 7.99 |
| SERPENT | 7.99 |
| Twofish | 7.99 |
| RC6 | 7.99 |

Figure 3.14 shows a three-dimensional representation of a ciphertext output generated by the SEPAR algorithm.

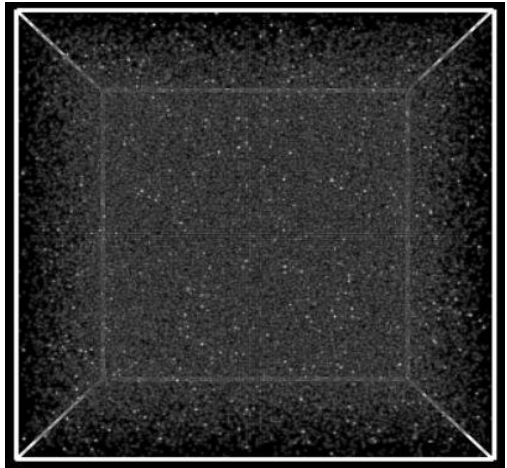

Fig 3.14: 3D representation of SEPAR output

### 3.2.9 Implementation

This section presents the results of the software implementation of the algorithm and evaluates the performance of the algorithm on a wide range of microcontrollers.
The Internet of Things (IoT) system can be classified into two sections: hardware and service-platform. In the hardware section, devices and equipment are categorized, which include wearable devices and embedded system-based equipment. Embedded system-based equipment is based on microcontrollers or microprocessors, and in these devices, development boards and tools are used. Various criteria, such as memory capacity, power consumption, purchase cost, execution speed, and the ability to support sensors and actuators by their microcontrollers and microprocessors, are considered when selecting development boards and tools. Table 3.9 provides a list of microcontrollers and microprocessors used, along with their applications.

Table 3-9: Widely adopted microcontrollers and microprocessors used in IoT development boards.

| Microcontroller/Microprocessor | Application |
|---|---|
| 8-Bit Atmega128 AVR Microcontroller from ATMEL family | Application in sensors, real-time controllers and Arduino boards |
| 16-bit MSPF430F1611 Microcontrllers family from Texas instruments | Peripherals and communication equipment |
| 32-bit ARM Microprocessor | Arduino boards and Raspberry Pi boards |

The goal of designing the algorithm is to achieve suitable performance on microcontrollers, microprocessors, development boards, development tools, and other equipment with limited resources. The measurable parameters for comparing performance are the number of execution cycles and throughput. Throughput is used as a performance metric for cryptographic algorithms and indicates the amount of data encrypted during the execution time of the algorithm. The throughput of a cryptographic algorithm is calculated using the following relation (3.13):

$$Throughput = \frac{Block\ size(bits)}{Execution\ Time(ms)} \qquad (3\text{-}13)$$

To clearly and transparently present the implementation details of the algorithm, the function call graph of the Shield algorithm for its implementation is provided. This graph is shown in Figure 3.15.

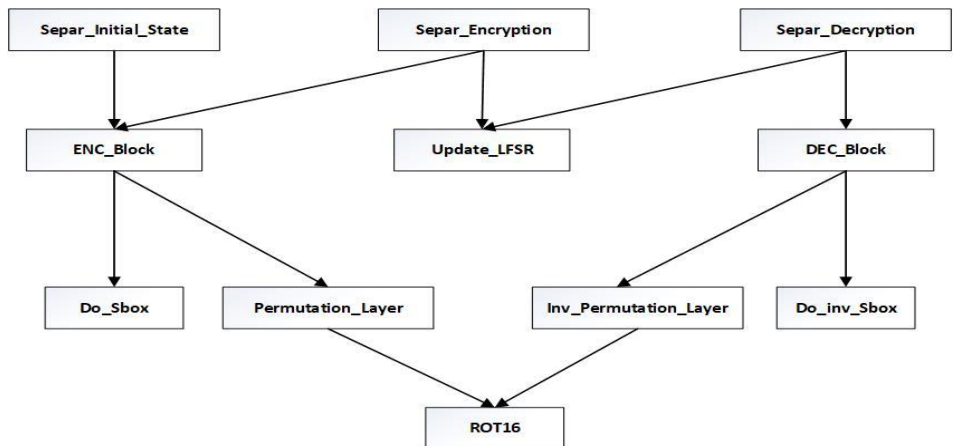

Figure 3-15: Function call graph used in the implementation

Next, the implementation of the algorithm on the microcontrollers listed in Table 3.9 and based on this function call graph will be presented.

### 3.2.10 Software Implementation on an 8-bit Microcontroller

The 8-bit ATmega128L microcontroller is a high-performance, low-power microcontroller widely used in applications requiring low energy consumption and long battery life. The implementation of the proposed algorithm on this microcontroller is presented below.

### 3.2.10.1  8-bit ATmega128L Microcontroller and Development Tools

The key features and practical applications of the 8-bit ATmega128L microcontroller are presented in Table 3.10.

Table 3-10: 8-bit ATmega128-L Microcontroller Features

| Risk guideline set architecture | 131 simple, optimized, and often single-cycle instructions |
|---|---|
| Memory | Programmable Flash Memory: 128KB |
|  | Internal SRAM Memory: 8KB |
| Clock Frequency | 0 to 8 KHZ |
| Power Source Supply | l2.7 to 5.5 volts |

To implement and test the efficiency of the SEPAR algorithm on the target platform, the Atmel Studio 7 Integrated Development Environment (IDE) was used for editing, simulation, and compilation.

### 3.2.10.2  Software Implementation on 8-Bit Atmega128L Microcontroller

To optimize the algorithm for memory usage, six subkeys generated for each Enc-block are created on-the-fly, so the memory required to store all the subkeys is not needed. The S-boxes are implemented as a byte array with 16 elements, where one nibble stores the substitution box values and the other nibble is filled with zeros. The S-boxes search for a 16-bit data block during encryption and decryption processes is done sequentially, processing 4 bits at a time. Additionally, to create memory-efficient code for this microcontroller, the compiler optimization level is set to "OPT=s".

### 3.2.10.3 Implementation Results

In Table 3-11, the memory consumption and number of machine cycles consumed for the shield algorithm and several lightweight algorithms on an 8-bit microcontroller at 4 MHz are shown with implementation.

Table 3-11: Memory Consumption and Cycle Count results on the 8-bit Microcontroller

| Cipher | Key size [*bit*] | Block size [*bit*] | 8-bit Microcontroller | Code size [*Kbytes*] | SRAM size [*bytes*] | int. [*cycles*] | Enc. [*cycles*] | Dec. [*cycles*] |
|---|---|---|---|---|---|---|---|---|
| SEPAR | 256 | 16 | *ATmega128L* | 3.86 | 32 | 39.154 | 9.761 | **9.783** |
| Present[3] | 80 | 64 | *ATmega163* | 2.39 | 32 | – | 646.166 | **634.616** |
| DESXL[30] | 184 | 64 | N/A | 3.19 | 0 | – | 8.531 | **7.961** |
| Salsa20[30] | 128 | 512 | N/A | 1.45 | 280 | – | 18.400 | **N/A** |
| HIGHT[30] | 128 | 64 | N/A | 5.67 | 0 | – | 2.964 | **2.964** |
| Hummingbird[4] | 256 | 16 | *ATmega128L* | 3.68 | 0 | 14.735 | 3.664 | **3.868** |

The results of this table indicate that the code size of the SEPAR algorithm is lower compared to the HITE algorithm, although the HIGHT algorithm requires fewer cycles than the SEPAR algorithm. This difference can be attributed to the block and stream-based combined structure of the SEPAR algorithm, which requires a preparation process. Additionally, the SEPAR algorithm necessitates a longer preparation time compared to the Hummingburd algorithm due to the greater number of internal states. Furthermore, a significant difference in the cycle counts between the SEPAR algorithm and the Present algorithm is observed, which can be attributed to its combined structure.

Next, a comparison of the throughput of these ciphers is presented in Figure 3.16.

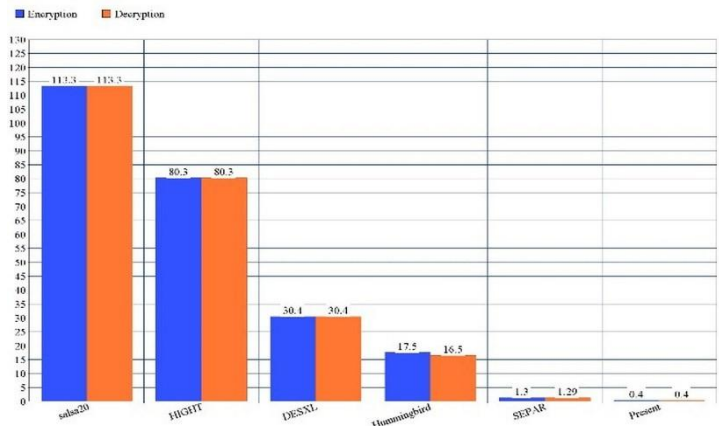

**Figure 3-16:** The overall throughput comparison of existing ciphers on 8-bit microcontroller

The results of this comparison show that SEPAR algorithm is 3 times faster than the Present algorithm but is 10 times slower than the Hummingburd algorithm. The main reason for this difference lies in the combined structure of the SEPAR algorithm. Additionally, it should be noted that SEPAR has a higher setup time compared to Hummingbird, which is due to having almost twice as many internal states. This increased number of states ensures the resilience of SEPAR against attacks that Hummingburd is vulnerable to. Tables 3.12 and 3.13 compare and evaluate the encryption and decryption execution times of the SEPAR algorithm with different message lengths.

Table 3-12: Encryption execution time of SEPAR in three different data-length at 4 MHz

| Message Length | **Present**-enc [3] (ms) | **Hummingbird**-enc[4] (ms) | **SEPAR**-enc (ms) | **DESXL**-enc (ms) | **HIGHT**-enc (ms) |
|---|---|---|---|---|---|
| 64-bit | 161.54 | 7.35 | 19.53 | 2.10 | 0.79 |
| 128-bit | 323.08 | 11.01 | 29.32 | – | – |
| 192-bit | 484.62 | 14.68 | 39.15 | – | – |

Table 3-13: Decryption execution time of SEPAR in three different data-length at 4 MHz

| Message Length | **Present**-dec [3] (ms) | **Hummingbird**-dec[4] (ms) | **SEPAR**-dec (ms) | **DESXL**-dec (ms) | **HIGHT**-dec (ms) |
|---|---|---|---|---|---|
| 64-bit | 158.65 | 7.55 | 19.59 | 2.10 | 0.79 |
| 128-bit | 317.31 | 11.42 | 29.53 | – | – |
| 192-bit | 475.96 | 15.29 | 39.29 | – | – |

## 3.2.11 ARM7 LPC2129 32-bit Microcontroller and Development Tools

The Arduino and Raspberry Pi boards are commonly used on the Internet of Things industry. These boards employ microprocessors with high computational power while still maintaining the characteristics of environments with limited resources. To evaluate the performance of the SEPAR algorithm in a 32-bit environment, the 32-bit ARM7 LPC2129 microcontroller was chosen.

### 3.2.11.1 Implementation on 32-Bit ARM LPC2129 processor and Development Tools

To evaluate the performance of the SEPAR algorithm in a 32-bit environment, the ARM7-based 32-bit LPC2129 microcontroller from NXP was selected. The key features of this microcontroller are shown in Table 3.14.

Table 3-14: LPC 2129 processor with ARM7 core key features

| CPU | ARM7 32-bit |
|---|---|
| Memory | Programmable Flash Memory: 128/256KB Internal SRAM Memory: 16KB |
| Clock Frequency | Maximum 60 MHz |
| Power Supply | 1.95 to 1.65 voltes |

To implement the algorithm on this microcontroller, the Keil Version "5.24.2.0" toolchain was used, which includes an integrated development environment (IDE) with a standard C compiler, core simulator, and debugging tools. In the implementation process, different levels of optimization were applied to the SEPAR algorithm to generate code with either low memory usage or high-speed characteristics.

### 3.2.11.2  32-bit platform implementation

To evaluate the performance of the SEPAR algorithm in a 32-bit environment, the 32-bit ARM7-based LPC2129 microcontroller from NXP was selected. For simulation on this processor, the Keil Version "5.24.2.0" toolchain was used, which includes an IDE with a standard C compiler, core simulator, and debugging tools. Additionally, various levels of optimization were applied to the SEPAR algorithm for generating code with either low memory or high-speed features.

### 3.2.11.3  Implementation Results

Table 3.15 [24] shows the number of machine cycles consumed for the SEPAR algorithm and several lightweight algorithms on the 32-bit processor.

Table 3-15: cycle count of well-known ciphers on 32-bit software platform at 12 MHZ

| **Cipher** | Structure | Key size [*bit*] | Block size [*bit*] | Number of cycles [*cycles*] |
|---|---|---|---|---|
| LED | SP-net | 128 | 64 | **425572.00** |
| KLEIN | SP-net | 96 | 64 | **10650.12** |
| **SEPAR** | **SP-net** | **256** | **16** | **2821.21** |
| BORON | SP-net | 128 | 64 | **7997.52** |
| Hummingbird2 | SP-net | 128 | 16 | **379812** |
| Present | SP-net | 64 | 128 | **3798.12** |
| Speck | Feistel | 128 | 64 | **588.24** |
| Simon | Feistel | 128 | 64 | **1268.04** |
| PICCOLO | Feistel | 128 | 64 | **2732.16** |
| CLEFIA | Feistel | 128 | 128 | **12576.12** |
| TWINE | Feistel | 128 | 64 | **7114.44** |

The number of encryption cycles for the SEPAR algorithm is the sum of two numbers: 11256 represents the cycles required for the setup phase, and 2821.21 represents the cycles required for the encryption operation. It's important to note that the setup phase is only performed once before encryption, so the number of cycles for encryption can be considered as 2821.21, which is less than the encryption cycles of other standard algorithms.
Additionally, a comparison of memory usage between the well-known algorithms and the SEPAR algorithm is shown in the graph in Figure 3.17 [24]

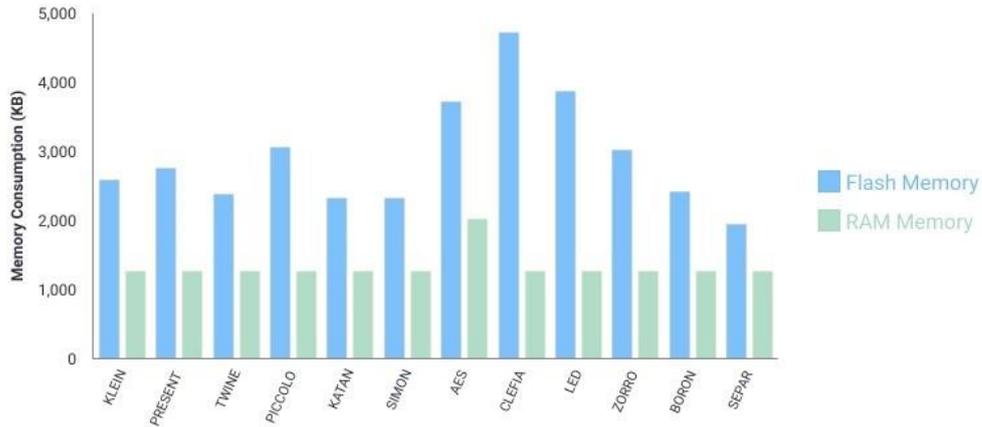
**Fig 3-17: Memory consumption in Lightweight cihphers**

In the graph shown in Figure 3.17, all algorithms are programmed using Embedded C and implemented on a 32-bit processor. This graph indicates that the flash memory required for SEPAR is 2220 bytes, and the RAM memory required is 1256 bytes. It demonstrates that SEPAR requires less memory compared to other standard algorithms.

**Table 3-16: Execution time of SEPAR for three different data-lengths at 12 MHz**

| Message length | Execution time ($\mu s$) |
|---|---|
| 64-bit | 195.4 |
| 128-bit | 289.7 |
| 192-bit | 384.3 |

To further analyze the performance of SEPAR, the encryption speed of the algorithm for three different message lengths—64, 128, and 192 bits—was calculated and is presented in Table 3.16

In Table 3.17, the results for throughput and encryption time are presented for commonly used lightweight algorithms, including SEPAR [24]. The throughput is calculated at a frequency of 12 MHz for the software platform .

**Table 3-17: Results of throughput and execution time at 12 MHz**

| Cipher | Key size [$bit$] | Block size [$bit$] | Execution time [$\mu s$] | Throughput [$kb/s$] |
|---|---|---|---|---|
| LED | 128 | 64 | 7092.86 | **9.00** |
| KLEIN | 96 | 64 | 887.51 | **72.00** |
| BORON | 128 | 64 | 666.46 | **96.02** |
| Hummingbird2 | 128 | 16 | 316.51 | **51.00** |
| **SEPAR** | **256** | **16** | **117.308** | **136.3** |
| Present | 64 | 128 | 2648.65 | 24.16 |
| Speck | 128 | 64 | 49.02 | **1305.00** |
| Simon | 128 | 64 | 105.67 | **605.00** |
| PICCOLO | 128 | 64 | 227.68 | **281.00** |
| CLEFIA | 128 | 128 | 1048.01 | 122.00 |
| TWINE | 128 | 64 | 592.87 | 108.00 |

## 3.2.12 Software Implementation on 16-bit mirocontroller

In the world of communications and telecommunications, 16-bit microcontrollers play a crucial and fundamental role due to their shared features that combine the advantages of both 8-bit and 32-bit microcontrollers. Figure 3.18 presents a Venn diagram comparing the features of both microcontroller types.

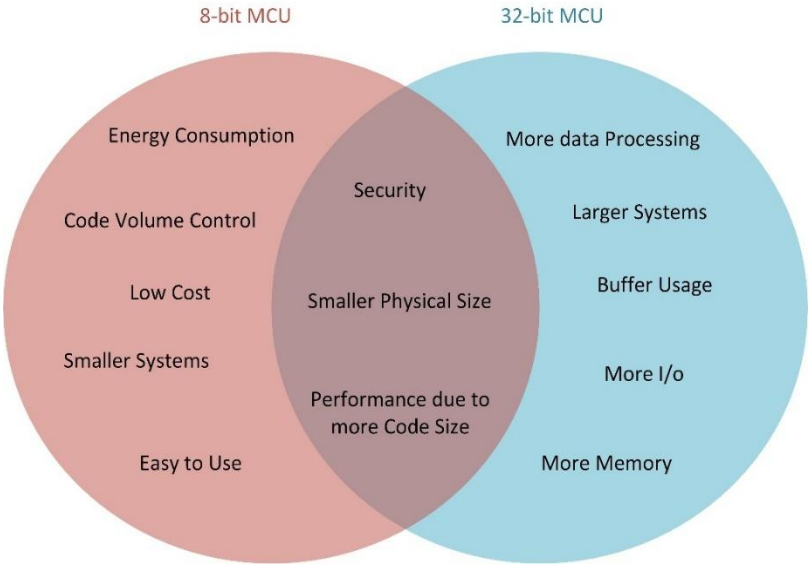

**Fi 3-18: Shared Features of the 16-bit microcontroller**

The MSP430 microcontroller family from Texas Instruments has widespread applications in the embedded systems industry due to its low power consumption. The MSP430-F1611 microcontroller has been selected as a reference microcontroller from this family for implementation. The next subsection presents the implementation of the proposed algorithm on this microcontroller.

### 3.2.12.1  MSP430 16-bit Microcontroller and Development Tools

To evaluate the performance of the SEPAR algorithm in a 16-bit platform, the MSP430-F1611 16-bit microcontroller has been selected. Due to the block structure and 16-bit internal variables in the SEPAR algorithm, the results of this implementation can be significant. The key features and applications of the MSP430-F1611 microcontroller are shown in Table 3.18.

Table 3-18: MSP430-F1611 microcontroller Key Features

| Von Neumann Architecture | All registers, RAM, flash, and peripherals use a common bus for data and addresses. |
|---|---|
| Instruction set Architecture | Includes 27 instructions and 7 different addressing modes that provide flexibility in data manipulation. |
| Memory | Programmable Flash Memory: 48 KB<br>Internal SRAM Memory: 10KB |
| Clock Frequency | 0 to 8 MHz |
| Power Supply | 1.8 to 3.6 volts |

To implement and simulate on this microcontroller, the IAR Embedded Workbench tool has been used, which includes a standard C language compiler, core simulator, and an integrated development environment. Additionally, different optimization levels have been applied to generate code with either low memory usage or high speed for the SEPAR algorithm.

### 3.2.12.2   Implementation on a 16-bit Platform

For the implementation of the algorithm, four subkeys for each block are generated on-the-fly, which eliminates the need for 64-bit memory to store these subkeys. In the 8-bit microcontroller, S-box is implemented as a one-byte array with 16 elements, and the S-box search process for a 16-bit data block is performed sequentially for both encryption and decryption algorithms. Additionally, to create memory-efficient code, the optimization level parameter is set to "Minimize Size."

### 3.2.12.3   Implementation Results

Table 3.19 shows the memory consumption, and the number of machine cycles consumed for the SEPAR algorithm and several lightweight algorithms on the 16-bit microcontroller.

Table 16: Memory Consumption and Cycle Count results at 4 MHZ on 16-bit Microcontroller

| Cipher | Key size [*bit*] | Block size [*bit*] | 16-bit Microcontroller | Code size [*Kbytes*] | Int. [*cycles*] | ENC. [*cycles*] | DEC. [*cycles*] |
|---|---|---|---|---|---|---|---|
| SEPAR | 256 | 16 | **MSP430F1611** | 4.67 | 31,127 | 7,845 | 7,849 |
| Present[3] | 80 | 64 | **C167CR** | 9.67 | – | 1,422,565 | 1,332,062 |
| Hummingbird[4] | 256 | 16 | **MSP430F1611** | 2.95 | 9,667 | 2,414 | 2,650 |

From the above table, it can be concluded that the Hummingbird algorithm consumes 78% less memory compared to the SEPAR algorithm. This difference is due to the number of internal states used in the SEPAR algorithm, which is almost twice that of the Hummingburd algorithm. Hummingbird has attempted to reduce memory consumption by decreasing the number of states in its structure, but this imbalance between cost and security has made the algorithm vulnerable to divide-and-conquer attacks. To counter this attack, we have used almost twice the number of internal states in the algorithm to maintain a balance without compromising the algorithm's resistance. Although this balance results in higher cycle consumption in the SEPAR algorithm, it still uses fewer cycles than the Present algorithm. Another reason for the lower cycle count is the hybrid structure of the proposed algorithm and its 16-bit internal operations. After the initialization process, the SEPAR algorithm achieves throughput rates of 1.71 and 1.70 kilobits per second for encryption and decryption operations, respectively. Figure 3.19 shows a comparison of the throughput of the SEPAR algorithm with the two lightweight algorithms, Present and Hummingbird.

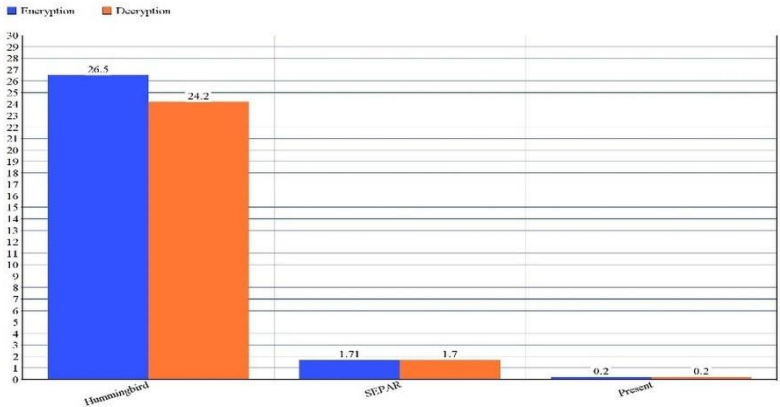

**Fig. 3.19: The Overall throughput comparison at 4 MHZ (on 16-bit Microcontroller)**

Based on the overall throughput results shown in the figure above, the SEPAR algorithm is 9 times faster than Present but 14 times slower than Hummingbird. The reason for this speed increase compared to Present can be attributed to its hybrid structure. The SEPAR algorithm has a longer initialization phase than the Hummingburd algorithm, which is due to the higher number of internal states. The use of more internal states in SEPAR ensures its security against attacks that Hummingbird is vulnerable to. In other words, in the Hummingburd algorithm, the security-cost trade-off has not been properly balanced. Tables 3.20 and 3.21 show the execution time of the SEPAR algorithm for different message lengths at a frequency of 4 MHz .

**Table 3-20:**
**Encryption execution time of SEPAR in three different data-length at 4 MHz on 16-bit platform**

| Message Length | **Present**-enc [3] (ms) | **Hummingbird**-enc[4] (ms) | **SEPAR**-enc (ms) |
|---|---|---|---|
| 64-bit | 360.64 | 4.83 | 10.24 |
| 128-bit | 721.28 | 7.24 | 14.78 |
| 192-bit | 1081.92 | 9.66 | 21.53 |

**Table 3-21:**
**Decryption execution time of SEPAR in three different data-length at 4 MHz on 16-bit platform**

| Message Length | **Present**-dec [3] (ms) | **Hummingbird**-dec[4] (ms) | **SEPAR**-dec (ms) |
|---|---|---|---|
| 64-bit | 333.02 | 5.07 | 10.91 |
| 128-bit | 666.03 | 7.72 | 15.02 |
| 192-bit | 1081.92 | 9.66 | 21.70 |

### 3.2.13  Other Software Implementations

For practical studies and educational purposes, a framework named TARA has been developed by the Cryptography and Security Research Center. Figure 3.20 illustrates the environment of the SEPAR and its plugin in this framework. Additionally, for further security analysis and evaluations, the Python implementation of the SEPAR algorithm is also available.

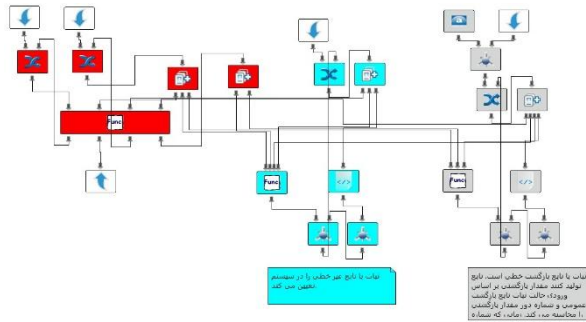

**Figure 3-20: SEPAR implementation on TARA framework**

## Summary and Conclusion

In this section, the new encryption algorithm called SEPAR, which combines block cipher and stream cipher methods, was presented. The overall structure of the algorithm both schematically and in pseudocode is presented. In addition, its implementation was carried out on software platforms for 8-bit, 16-bit, and 32-bit microcontrollers. In the next chapter, the conclusion and performance improvement of the SEPAR algorithm compared to other algorithms mentioned will be provided.

# 4) Chapter Four: Conclusion and Recommendations

**Chapter Objectives:**

- Conclusion

- Future Work Recommendations

# Introduction

In this thesis, a novel hybrid encryption scheme named *SEPAR* has been presented. The hybrid design approach, which combines block cipher and stream cipher techniques, enhances encryption speed while simultaneously improving the algorithm's resistance against various cryptographic attacks. In this chapter, the performance improvement of the SEPAR algorithm is evaluated across 8-bit, 16-bit, and 32-bit software platforms using different message lengths.

## 4.1 Performance Improvement Analysis on 8-bit Microcontrollers

The performance improvement of the algorithm on 8-bit microcontrollers is presented in Tables 4-1 and 4-2.

**Table 4-1:**
**The overall SEPAR Encryption Performance improvement at 4 MHZ on 8-bit Microcontroller**

| Message Length | Performance Improvement to **Present** | Performance improvement to **Hummingbird** | Performance improvement to **DESXL** | Performance improvement to **HIGHT** |
|---|---|---|---|---|
| 64-bit | 87.91% | -165% | Less than -200% | Less than -200% |
| 128-bit | 90.92% | -166% | — | — |
| 192-bit | 91.92% | -166% | — | — |

**Table 4-2:**
**The overall SEPAR Decryption Performance improvement at 4 MHZ on 8-bit Microcontroller**

| Message Length | Performance Improvement to **Present** | Performance improvement to **Hummingbird** | Performance improvement to **DESXL** | Performance improvement to **HIGHT** |
|---|---|---|---|---|
| 64-bit | 87.65% | -159% | Less than -200% | Less than -200% |
| 128-bit | 90.69% | -159% | — | — |
| 192-bit | 91.75% | -159% | — | — |

Message lengths of 64, 128, and 192 bits are utilized. In Figure 4-1, the SEPAR algorithm is compared with other lightweight encryption algorithms. It can be concluded that the design of SEPAR achieves a balanced trade-off between security and performance .

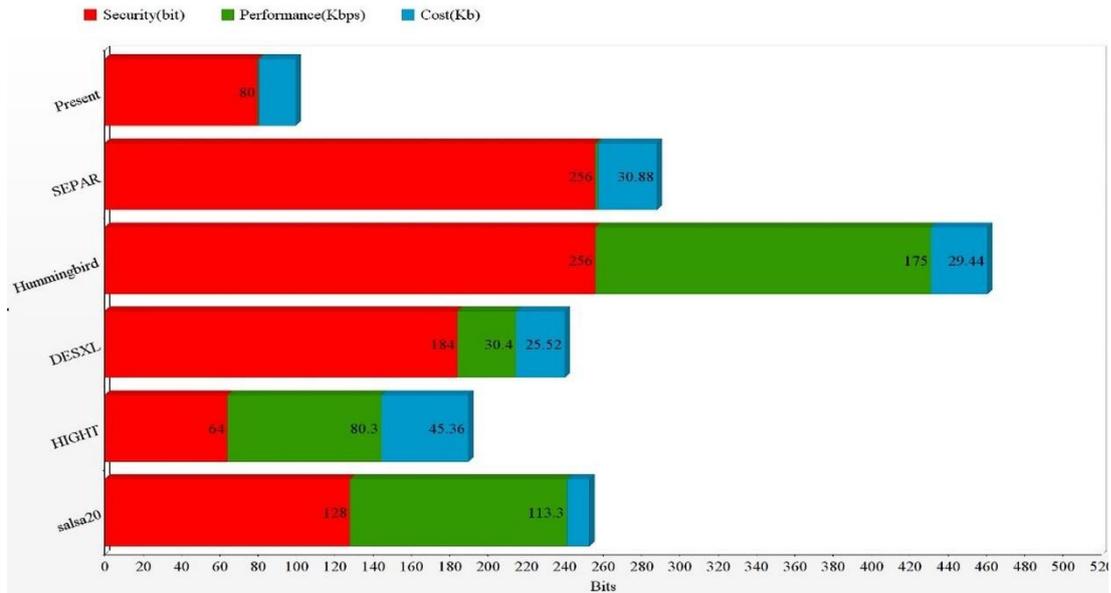
**Fig. 4-1: Comparison based on design goals of security, cost, and performance on 8-bit microcontroller**

The above comparison indicates that although SEPAR utilizes a 256-bit key, it does not offer optimal performance and consumes more memory in comparison.

## 4.2 Performance Improvement Analysis on a 16-bit Microcontroller

The analysis of the performance improvement of the SEPAR algorithm on a 16-bit microcontroller is presented in Tables 4-3 and 4-4.

**Table 24:**
**The overall SEPAR Encryption Performance improvement at 4 MHZ on 16-bit Microcontroller**

| Message Length | Performance Improvement to **Present** | Performance improvement to **Hummingbird** |
|---|---|---|
| 64-bit | 97.16% | -112% |
| 128-bit | 97.95% | -41.43% |
| 192-bit | 98.01% | -122% |

**Table 25:**
**The overall SEPAR Decryption Performance improvement at 4 MHZ on 16-bit Microcontroller**

| Message Length | Performance Improvement to **Present** | Performance improvement to **Hummingbird** |
|---|---|---|
| 64-bit | 96.72% | -115% |
| 128-bit | 97.74% | -94.55% |
| 192-bit | 97.99% | -124% |

The data presented in these tables indicate that SEPAR provides approximate performance improvement ranging from 96.72% to 98.01% when it is employed instead of PRESENT for encryption and decryption operations across message lengths of 64, 128, and 192 bits. In Figure 4-2, the SEPAR algorithm is compared with other lightweight algorithms.

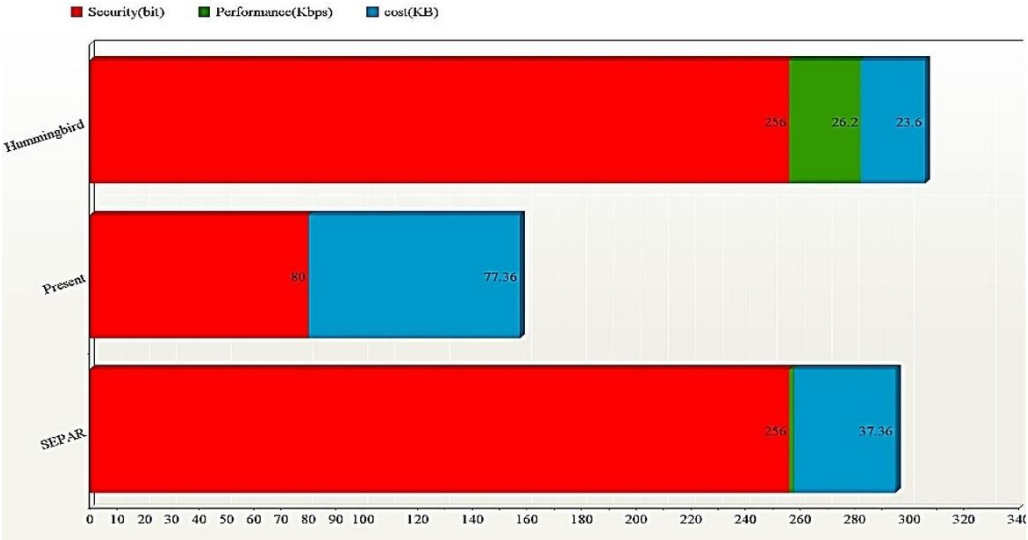

**Fig. 4-2: Comparison based on design goals of security, cost, and performance on 16-bit microcontroller**

As illustrated in the above figure, the SEPAR design better meets the objectives of security, performance, and cost compared to PRESENT. This is due to SEPAR offering a longer key length and higher performance while consuming less memory. However, in comparison to the Hummingbird algorithm, SEPAR does not perform as well. Although Hummingbird has the same key length as SEPAR, it delivers higher efficiency and incurs lower implementation costs.

## 4.3 Evaluation of Performance Improvement on a 32-bit Microcontroller

The assessment of SEPAR's performance improvement on a 32-bit microcontroller is presented in Table 4-5.

**Table 26:**
**The overall SEPAR Encryption Performance improvement at 12 MHZ on 32-bit ARM processor**

| SEPAR Performance improvement to | Computed Performance improvement |
|---|---|
| Present | 464.15% |
| KLEIN | 88.93% |
| HUMMINGBIRD2 | 16.72% |
| BORON | 42.22% |

The table demonstrates that SEPAR offers a significant performance improvement compared to well-known lightweight ciphers such as PRESENT, CLEFIA, and BORON when implemented on a 32-bit software-based platform. In Figure 4-3, SEPAR is compared with the existing lightweight algorithms. Based on the results of this comparison, it can be concluded that the SEPAR design provides an appropriate balance between security, cost, and performance on a 32-bit ARM processor.

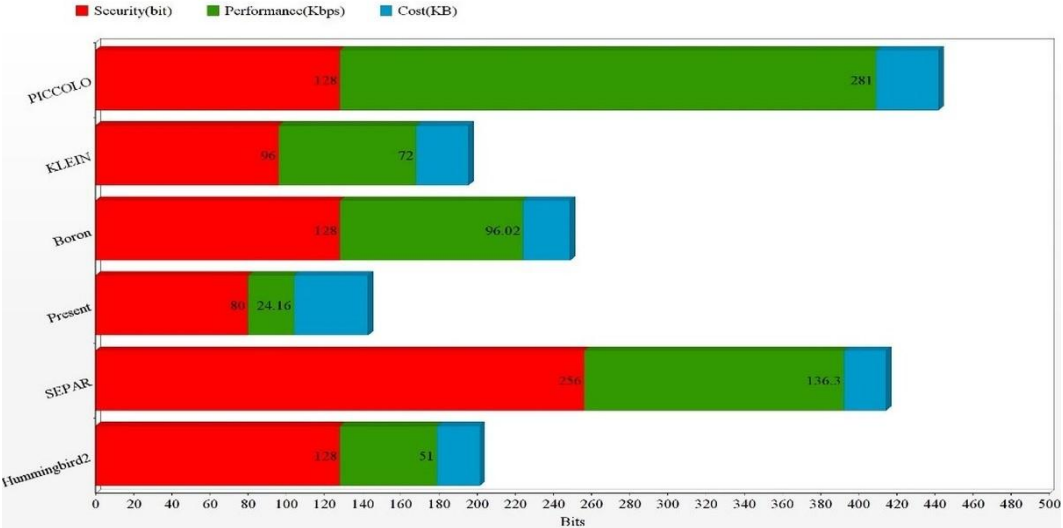

**Fig. 4-3: Comparison of existing ciphers based on design goals of security, cost, and performance**

As illustrated in the figure above, the design of SEPAR achieves a simultaneous balance among key design objectives. In other words, by employing a 256-bit key and maintaining low memory consumption, SEPAR offers better performance compared to other lightweight algorithms on a 32-bit ARM processor.

# 5 Further work

The SEPAR cryptographic design, which follows a hybrid architecture, has been implemented on 8-bit, 16-bit, and 32-bit software platforms. Its security has also been evaluated against various cryptographic attacks. In future research, the algorithm's resistance can be analyzed against more advanced attacks, including implementation attacks.